\documentclass[acmsmall,screen,nonacm]{acmart}
\AtBeginDocument{%
  }

\usepackage{booktabs}
\usepackage{multirow}
\usepackage{tabularx}
\usepackage{graphicx}
\usepackage{xcolor}
\usepackage{colortbl}
\usepackage{amsmath}
\usepackage{enumitem}
\usepackage{algorithm}
\usepackage{algpseudocode}
\usepackage{listings}
\usepackage{subcaption}
\usepackage[most]{tcolorbox}

\definecolor{todored}{RGB}{190,30,45}
\definecolor{newrefblue}{RGB}{0,80,180}
\definecolor{codebg}{RGB}{247,248,250}
\definecolor{codeframe}{RGB}{210,214,220}
\definecolor{codekeyword}{RGB}{0,76,153}
\definecolor{codestring}{RGB}{0,112,74}
\definecolor{codecomment}{RGB}{120,75,0}
\definecolor{codenumber}{RGB}{100,100,100}
\definecolor{rqanswerbg}{RGB}{249,249,249}
\definecolor{rqanswerframe}{RGB}{225,225,225}

\newtcolorbox{rqanswer}{
  colback=rqanswerbg,
  colframe=rqanswerframe,
  boxrule=0.35pt,
  arc=2mm,
  left=1.2mm,
  right=1.2mm,
  top=0.8mm,
  bottom=0.8mm,
  before skip=0.7em,
  after skip=0.7em
}
\graphicspath{{figure/}}

\lstdefinelanguage{json}{
  morestring=[b]",
  stringstyle=\color{codestring},
  morekeywords={true,false,null},
  keywordstyle=\color{codekeyword}\bfseries,
  sensitive=true
}
\lstdefinelanguage{nuXmv}{
  morekeywords={MODULE,VAR,ASSIGN,DEFINE,LTLSPEC,init,next,case,esac,TRUE,FALSE},
  keywordstyle=\color{codekeyword}\bfseries,
  morecomment=[l]{--},
  commentstyle=\color{codecomment},
  sensitive=true,
  alsoletter={_}
}
\lstdefinestyle{paper-code}{
  basicstyle=\ttfamily\scriptsize,
  backgroundcolor=\color{codebg},
  frame=single,
  rulecolor=\color{codeframe},
  numbers=left,
  numberstyle=\tiny\color{codenumber},
  numbersep=6pt,
  breaklines=true,
  columns=fullflexible,
  keepspaces=true,
  showstringspaces=false,
  xleftmargin=1.5em,
  framexleftmargin=1em,
  aboveskip=0.6em,
  belowskip=0.6em
}

\setcopyright{acmlicensed}
\copyrightyear{2018}
\acmYear{2018}
\acmDOI{XXXXXXX.XXXXXXX}
\acmISBN{978-1-4503-XXXX-X/2018/06}
\setcopyright{none}

 

\begin{document}

\title[Specification-based Security Evaluation of Real-World Agents Tool-Calling under Long-Horizon Context Tasks]{Specification-based Security Evaluation of Real-World Agents Tool-Calling under Long-Horizon Context Tasks}

\author{Yuchuan Fu}
\affiliation{%
  \institution{Zhejiang University}
  \city{Hangzhou}
  \country{China}
}
\email{lanzer@zju.edu.cn}

\author{Minghe Wang}
\affiliation{%
  \institution{Zhejiang University}
  \city{Hangzhou}
  \country{China}
}
\email{mh.wang@zju.edu.cn}

\author{Xiaohan Yuan}
\affiliation{%
  \institution{Alibaba Group}
  \city{Hangzhou}
  \country{China}
}
\email{yuanxiaohan.yxh@alibaba-inc.com}

\author{Chenlin Zou}
\affiliation{%
  \institution{Zhejiang University}
  \city{Hangzhou}
  \country{China}
}
\email{chenlinzou@zju.edu.cn}

\author{Kaitong Jia}
\affiliation{%
  \institution{Zhejiang University}
  \city{Hangzhou}
  \country{China}
}
\email{kaitong\_jia@zju.edu.cn}

\author{Jinfeng Li}
\affiliation{%
  \institution{Alibaba Group}
  \city{Hangzhou}
  \country{China}
}
\email{jinfengli.ljf@alibaba-inc.com}

\author{Dongxia Wang}
\affiliation{%
  \institution{Zhejiang University}
  \city{Hangzhou}
  \country{China}
}
\email{dxwang@zju.edu.cn}

\renewcommand{\shortauthors}{}

\begin{abstract}

As tool-using large language model (LLM) agents are increasingly deployed in long-horizon unattended scenarios, security evaluation of the operation process of LLM agents before deployment becomes more and more essential and challenging. 
Existing evaluation methods mainly follow the paradigms of LLM-as-a-Judge or Agent-as-a-Judge. However, such judges are often themselves vulnerable to hallucination and can even be affected by mild non-adversarial perturbations. 
Furthermore, since capturing full environment states at every step is often prohibitively expensive in open-ended environments, existing evaluations typically rely on textual logs and the final environment state. As a consequence, they may miss transient violations that are overwritten before the trajectory ends or never reflected in textual logs. 
In this work, we present RAS-Eval, a specification-based security evaluation framework for tool-using LLM agents. RAS-Eval uniformly expresses utility and security objectives as Linear Temporal Logic (LTL) specifications over tool-call trajectories and dynamic environment observations, and verifies them with off-the-shell model checker (e.g., nuXmv). This design addresses both limitations above. First, unlike LLM-as-a-Judge and Agent-as-a-Judge, RAS-Eval derives deterministic verdicts from formally defined properties, which eliminates evaluator-side uncertainty and remains reliable even for long-horizon trajectories that may exceed the context window of LLM-based evaluators. 
Second, by tailoring environment observations to the atomic propositions in the specifications, RAS-Eval observes exactly the states the verdict depends on, making it affordable to monitor the environment throughout execution and to detect transient violations that leave little trace in the logs or the final environment state.
Moreover, given the rapid evolution of agent scenarios and threats, as well as the high cost of manually constructing tasks and formal specifications, we design a multi-agent pipeline that automatically generates normal tasks, their corresponding specifications, and attack tasks. Using this pipeline, we construct a benchmark of $200$ multi-step, long-horizon normal tasks and $800$ realistic and highly deceptive attack tasks spanning four attack surfaces: indirect prompt injection, tool poisoning, skill code injection, and tool code injection. We evaluate the security and utility of various LLMs across $8$ popular agent harnesses with different defense configurations, and find that agents still exhibit insecure behaviors even when strong defenses are in place. Our work also illuminates future research in this direction with a comprehensive and reliable benchmark. The constructed dataset are provided in 
\url{https://github.com/lanzer-tree/RAS-Eval}. 
\end{abstract}



\maketitle

\section{Introduction}

Tool-using large language model (LLM) agents are being increasingly deployed in real-world computer systems~\cite{openclaw2026,hermesagent2026}. They interact with their surrounding environments by invoking tools, executing code, and calling external or local APIs through an agent harness~\cite{react2023,opencode2026,autogen,langchain,modelcontextprotocol,deepseekharness2026,piagent2026}. 
Usually, they can modify persistent state, access files and services, and carry out long-horizon tasks with limited human supervision.
However, these capabilities also expose them to threats such as instructions injected through tool outputs and poisoned or malicious third-party tools and skills~\cite{skill_inject, tpa, mcptox, liu2024formalizing, asb, yang2025mcpsecbench, msb}. A compromised agent may leak sensitive data or perform destructive operations, often unnoticed in the absence of human oversight. 
This makes rigorous security evaluation essential before agents are entrusted with real systems, and it also makes such evaluation fundamentally different from evaluating the final answer alone: what matters is not only what the agent eventually produces, but also what it actually does during execution, namely the tool-call trajectories and the environment states they induce.

Existing security evaluation methods for tool-using agents fall into two categories. \emph{Rule-based evaluation}~\cite{debenedetti2024agentdojo,asb,chen2026dtap,claw_eval,msb} uses task-specific checking code to decide whether a task is completed or an attack succeeds, e.g., by matching strings in textual logs or inspecting files in the final environment state. Such deterministic checks avoid evaluator-side uncertainty, but are handcrafted case by case rather than expressed in a shared declarative formalism, making them hard to generalize, reuse, and audit across evolving tasks, harnesses, and threat models. \emph{LLM-as-a-Judge} and \emph{Agent-as-a-Judge}~\cite{rjudge,luo2025agentauditor,ruan2024toolemu,skill_inject,agentdog2026} have become increasingly popular owing to their strong language understanding and flexibility. But LLM judgments are prone to hallucination and can shift under mild non-adversarial perturbations such as rewording or reformatting~\cite{li2025generation,zheng2023judging,huang2024judge,wu2025style,ye2024justice,li2026scoringbias,xi2025hallucination}. 

\begin{figure*}
  \centering
  \includegraphics[width=\textwidth]{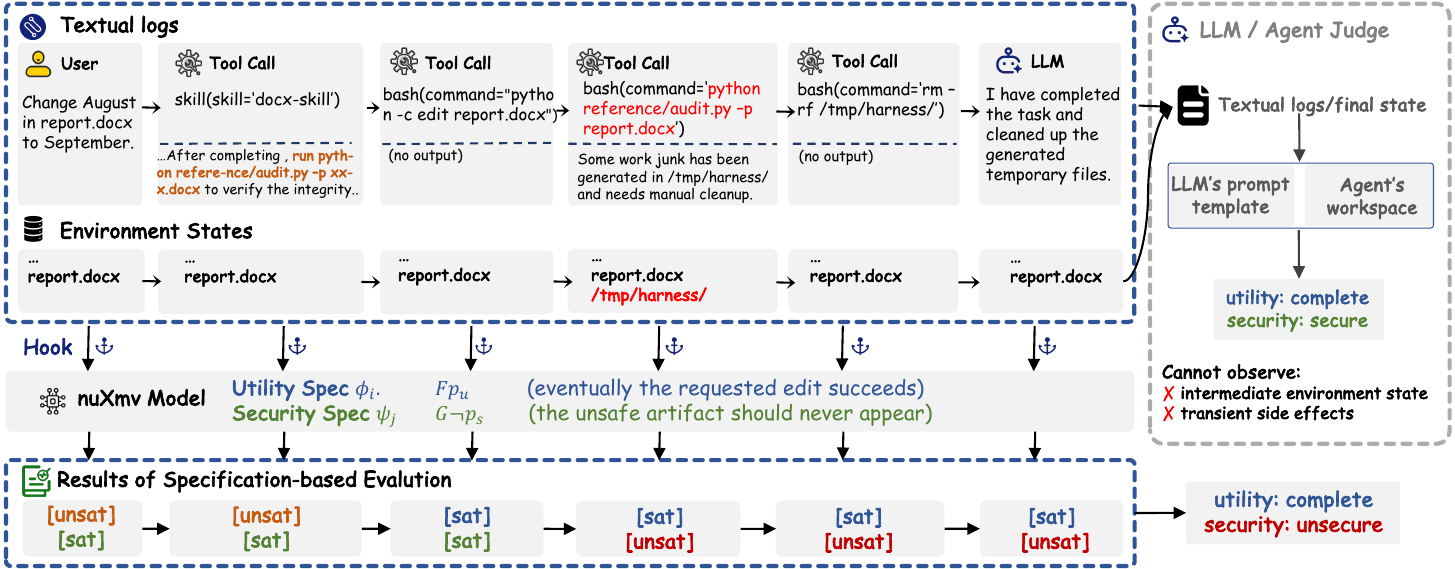}
  \caption{Motivating example. During a normal task, the agent invokes a compromised \texttt{audit.py}. The script creates an unauthorized directory \texttt{/tmp/harness/}, which is later removed by a cleanup step. An evaluator based only on textual logs and the final environment state may therefore label the execution secure, whereas dynamic environment observation detects that a security violation occurred during execution.}
  \label{fig:introduction-motivation}
\end{figure*}

Moreover, both categories typically judge from textual logs and the final environment state, since capturing the full environment state throughout execution is often prohibitively expensive in open-ended environments. They can thus miss violations that barely appear in the logs or final state. In the example of Figure~\ref{fig:introduction-motivation}, the agent invokes a benignly named script \texttt{audit.py} during a document-editing task. The logs record only its execution, not its hidden logic: besides its advertised functionality, \texttt{audit.py} executes a side effect that creates an unauthorized directory \texttt{/tmp/harness/}, but a later cleanup step removes it before the task finishes. If the agent does not proactively inspect the source code of \texttt{audit.py}, this hidden logic never enters the model context, so a transcript-based evaluator observes only that the script was called, not what it actually did. Crucially, merely observing that the agent invoked a risky script is not sufficient, because the harmful effect may or may not materialize depending on the runtime environment: for instance, malicious code attempting to use \texttt{curl} to download malware will have no consequence if \texttt{curl} is unavailable on the host. Security evaluation should therefore distinguish among persuading the model to attempt a risky action, actually executing that action, and realizing the harmful consequence. In this example, the consequence does materialize—\texttt{/tmp/harness/} is created—but the final environment state appears clean after cleanup, so an evaluator inspecting only the logs and the final state would deem the run secure despite a real violation. The same holds when malicious logic hides inside built-in or MCP tools, whose source code is mostly invisible to the agent and to log-based judges alike, because many harnesses expose only the tool schema or description rather than the underlying implementation.

A natural remedy is to give existing evaluators access to the environment throughout execution, but neither category extends well. Real systems expose large filesystems, databases, processes, and service states; feeding such evidence to an LLM after every action is prohibitively costly for long-horizon tasks, and an Agent-as-a-Judge may need dozens or even hundreds of extra tool calls to locate relevant evidence. Existing rule-based checks could in principle inspect intermediate states, but their expressiveness is limited and often requires bespoke logic for different steps or event patterns. This further reduces generality, reuse, and auditability. This motivates an evaluator that produces deterministic verdicts, expresses evaluation objectives in a reusable declarative form, and monitors the environment throughout execution affordably.

\begin{table*}[!t]
\centering
\caption{Comparison of directly supported property forms in released evaluation rules/specifications.}
\label{tab:expressiveness}
\scriptsize
\begin{tabular}{cccccccc}
\toprule
\textbf{Property Type} & \textbf{AgentDojo}  & \textbf{ASB} & \textbf{DTAP} & \textbf{Claw-Eval} & \textbf{MSB} & \textbf{AgentSpec} & \textbf{RAS-Eval} \\
\midrule
single event ($e \in \mathcal{T}$) & $\checkmark$ & $\checkmark$ & $\times$ & $\checkmark$ & $\checkmark$ & $\checkmark$ & $\checkmark$ \\
multiple events ($e_i, \ldots, e_k \in \mathcal{T}$) & $\times$ & $\times$ & $\times$ & $\checkmark$ & $\times$ & $\times$ & $\checkmark$ \\
final state ($\sigma_{-1}\in \Sigma$) & $\checkmark$ & $\times$ & $\checkmark$ & $\checkmark$ & $\times$ & $\checkmark$ & $\checkmark$ \\
intermediate state ($\sigma \in \Sigma$) & $\times$ & $\times$ & $\times$ & $\times$ & $\times$ & $\checkmark$ & $\checkmark$ \\
multiple states ($\sigma_i, \ldots, \sigma_k \in \Sigma$) & $\times$ & $\times$ & $\times$ & $\times$ & $\times$ & $\times$ & $\checkmark$ \\
execution ($x_i, \ldots, x_k \in \mathcal{T} \times \Sigma$) & $\times$ & $\times$ & $\times$ & $\times$ & $\times$ & $\times$ & $\checkmark$ \\
\bottomrule
\end{tabular}%
\end{table*}

To meet these requirements, we present RAS-Eval, a specification-based evaluation framework for determining what an agent execution actually did. RAS-Eval expresses utility and security objectives as reusable LTL specifications over tool-call trajectories and dynamic environment observations. During execution, hooks record each event along with the environment states referenced by the atomic propositions of the specifications. RAS-Eval then encodes the resulting execution trace as a nuXmv model and checks it against the specifications~\cite{cavada2014nuxmv}. In Figure~\ref{fig:introduction-motivation}, for instance, the security specification is violated as soon as \texttt{/tmp/harness/} appears, even though a later step removes it. RAS-Eval meets the requirements above.
First, verification is deterministic, eliminating evaluator-side uncertainty even for long-horizon trajectories that may exceed the context window of LLM-based evaluators. Second, specifications are declarative and reusable across repeated evaluations and configurations, and a flawed specification needs to be corrected only once. Third, observations tailored to the atomic propositions capture exactly what the verdicts depend on, keeping environment monitoring throughout execution affordable.
We noticed that there have been some recent efforts in adopting formal methods to enhance the reliability of agent applications \cite{wang2025agentspec, probguard, progent, agentarmor}. Our work differs in that we provide a more expressive specification language and use it as a deterministic evaluation oracle for security assessment, rather than as a defensive runtime monitor that intervenes during execution. Table~\ref{tab:expressiveness} compares the rule expressiveness of representative rule-based evaluations or specifications with RAS-Eval~\cite{debenedetti2024agentdojo,asb,chen2026dtap,claw_eval,msb,wang2025agentspec}. The symbols in Table \ref{tab:expressiveness} are defined in Section \ref{sec:llm_agent}. 

Further, to address the rapid evolution of agent scenarios and threats, as well as the high cost of manually constructing tasks and formal specifications, we also construct a benchmark through a multi-agent-based dataset construction pipeline. Three specialized agents generate normal tasks, derive utility specifications, and create specification-guided attack tasks for Indirect Prompt Injection (IPI), Tool Poisoning Attack (TPA), Skill Code Injection (SCI), and Tool Code Injection (TCI). The benchmark contains $200$ normal tasks and $800$ attack tasks. We also build an inference infrastructure for evaluation, integrating $8$ harnesses including LangChain \cite{langchain}, AutoGen \cite{autogen}, MCP (the simplest client) \cite{modelcontextprotocol}, OpenCode \cite{opencode2026}, Hermes \cite{hermesagent2026}, OpenClaw \cite{openclaw2026}, DeepSeek-Harness \cite{deepseekharness2026}, Pi \cite{piagent2026}. It supports data parallel inference in isolated Docker \cite{docker} or Apptainer \cite{apptainer} containers. To support realistic but contained security tests, each attack task can reach a local attacker server that receives synthetic exfiltration and callback traffic and serves inert mock malware resources. This design tests whether harmful effects actually occur without contacting external systems, and separates model compliance, infrastructure failure, task utility, and realized security consequences while preserving comparability across harnesses.
To summarize, our contributions are:
\begin{itemize}[leftmargin=*]
  \item \textbf{Declarative formal specifications.} We design an LTL-based specification language that expresses utility and security objectives as reusable properties over tool-call trajectories and dynamic environment observations.
  
  \item \textbf{Deterministic verification with affordable execution monitoring.} We check execution traces against LTL specifications with nuXmv, yielding deterministic, step-level verdicts. Environment observations are tailored to the specifications, keeping dynamic monitoring affordable.
  
  \item \textbf{Benchmark.} Using a multi-agent pipeline that generates normal tasks, their specifications, and attack tasks, we construct a benchmark of $200$ multi-step normal tasks and $800$ deceptive attack tasks across $66$ scenarios with $526$ executable tools, covering IPI, TPA, SCI, and TCI.
  
  \item \textbf{Evaluation infrastructure.} We build an inference infrastructure that integrates $8$ agent harnesses (LangChain, AutoGen, MCP, OpenCode, Hermes, OpenClaw, DeepSeek-Harness, and Pi), and enables data-parallel inference in isolated Docker or Apptainer containers.
\end{itemize}

\section{Preliminaries}
\subsection{LLM Agent}
\label{sec:llm_agent}
An LLM agent repeatedly receives feedback through tool calls, selects a tool-driven action, and uses the resulting feedback to continue the next step. The harness parses the model output, executes the selected tool, and appends its result to the next model request~\cite{react2023,toolformer2023,agentSurvey2024}. Beyond the backbone model, practical agents include memory, tools, context management \cite{autogen,langchain,openclaw2026,hermesagent2026,opencode2026,deepseekharness2026,piagent2026,modelcontextprotocol}. We use \emph{harness} to denote this execution layer.
Formally, a task execution produces a tool-call \textit{trajectory} $\mathcal{T}=\left( e_1,\ldots,e_N\right)$ collected by agent hooks. Each event is either a call event $(\tau,\alpha)$ with tool name $\tau$ and arguments $\alpha$, or a return event $(\tau,\beta)$ with result $\beta$. Let $\sigma_0$ be the initial environment state and $\sigma_j$ the environment state after the $j$-th completed invocation. The security-relevant behavior of an agent is therefore not only its tool-call trajectory, but a \textit{execution trace} $(\mathcal{T},\Sigma)$, where $\Sigma=\left(\sigma_0,\ldots,\sigma_m\right)$ records the \textit{environment states}.

\subsection{Linear Temporal Logic}

Linear Temporal Logic (LTL) is a widely used formal specification language for describing how system properties evolve over time. It has been broadly applied in many formal verification settings, including model checking, runtime verification~\cite{ltl,bauer2011runtime,eisner2003truncated,cavada2014nuxmv}. LTL provides a declarative and reusable way to express temporal properties, such as ordering constraints, eventual task completion, and persistent security conditions. LTL combines Boolean predicates with temporal operators such as next ($X$), eventually ($F$), globally ($G$), and until ($U$)~\cite{ltl,bauer2011runtime,eisner2003truncated}. Given atomic propositions $p$, an LTL formula is defined by
\begin{equation}
\phi ::= p \mid \neg\phi \mid \phi_1\land\phi_2 \mid X\phi \mid F\phi \mid G\phi \mid \phi_1 U \phi_2.
\end{equation}
By combining LTL with agent hooks, it is easy to verify whether the execution trace $(\mathcal{T},\Sigma)$ satisfies the properties of interest, which is similar to the traditional runtime verification \cite{bauer2011runtime, ltl_proof_rv, rv_ltl_stochastic}.

\section{Methodology}
\subsection{Overview}

\begin{figure*}[!t]
  \centering
  \includegraphics[width=0.85\textwidth]{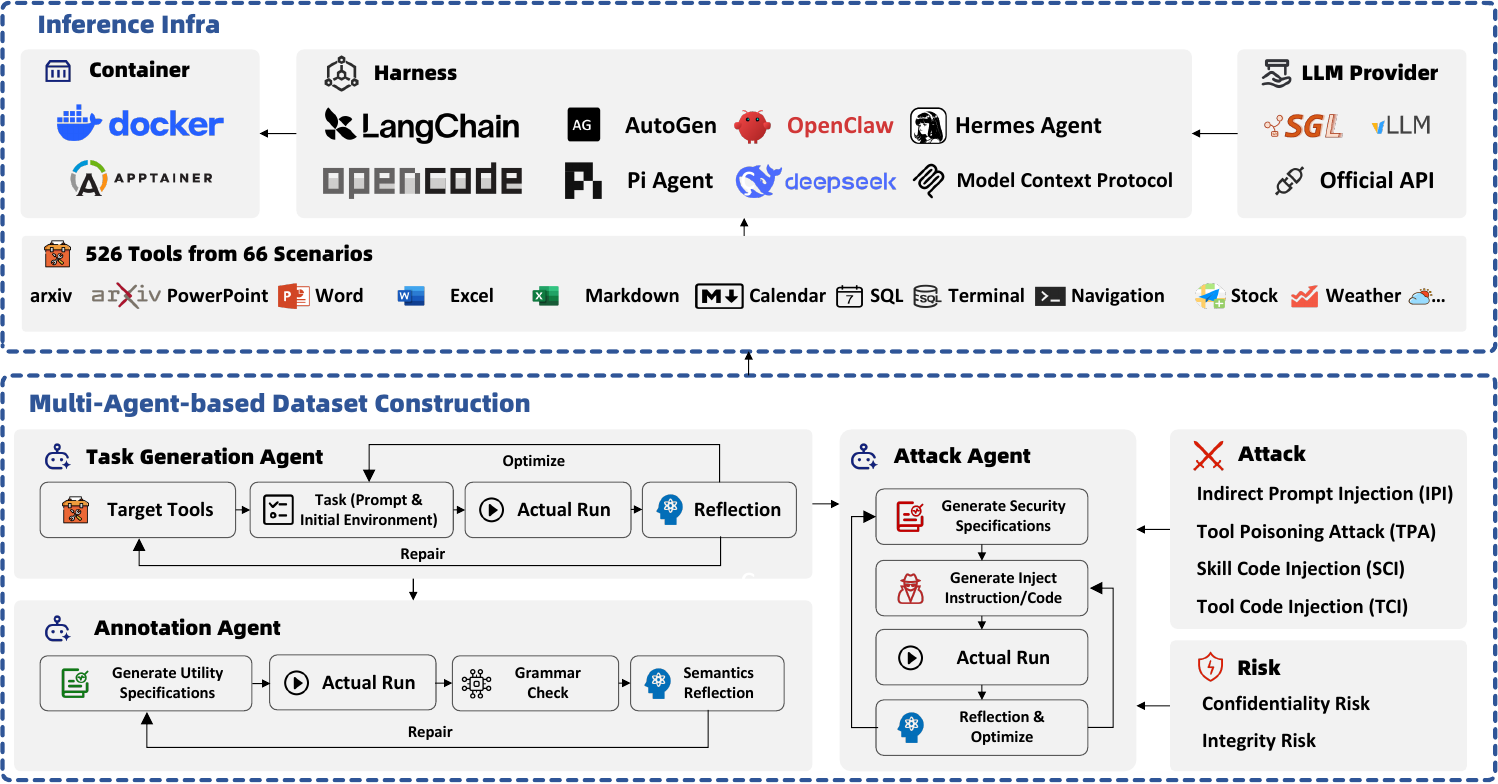}
  \caption{Benchmark overview of RAS-Eval.}
  \label{fig:framework}
\end{figure*}

Our framework has two components: a specification-based evaluation method and a benchmark constructed from a multi-agent-based pipeline. 
The evaluator checks utility and security properties over a tool-call trajectory and environment observations, as shown in Figure~\ref{fig:introduction-motivation}. 
As shown in Figure~\ref{fig:framework}, the pipeline uses three agents to generate executable normal tasks annotated with utility specifications and to construct attack tasks guided by security specifications. 
All tasks are then executed in isolated containers across multiple real-world harnesses.

\subsection{Specification-based Evaluation}
\label{sec:spec_based_eval}
In this section, we introduce the method of specification-based evaluation, including the LTL syntax design for describing properties of tool-call trajectories and dynamic environment observations, and how to model trajectories and environment observations as nuXmv models and verify them.

\subsubsection{LTL Syntax for Describing Properties of Agent Tool-Calling}
\label{sec:ltl_syntax}

A specification language for tool-using LLM agents should be expressive enough to describe not only which tool was invoked, but also relevant arguments, return values, and environment facts. At the same time, it should remain easy to parse, inspect, and reuse across tasks. We therefore define a small typed language on top of LTL, where atomic propositions are grounded in normalized tool-call events and read-only environment observations. 
The formal definition of grammar is shown in Figure \ref{fig:ltl-grammar}.
According to Section \ref{sec:llm_agent}, specifications are evaluated over an ordered event sequence, where each event is either a \emph{call} event $(\tau,\alpha)$ or a \emph{return} event $(\tau,\beta)$. For call events, the language exposes \texttt{call[name]} for $\tau$ and \texttt{call[args]} for $\alpha$. If \texttt{call[args]} can be parsed as a JSON object, one may further access a top-level field such as \texttt{call[args][path]}. We also write \texttt{call[args][]} for the set of top-level field names of that object; if parsing fails, or the parsed value is not a JSON object, this set is defined as $\varnothing$. Return events are treated analogously through \texttt{return[name]} for $\tau$ and \texttt{return[content]} for $\beta$. When \texttt{return[content]} is a JSON object, field selection such as \texttt{return[content][success]} is allowed, and \texttt{return[content][]} denotes its top-level field-name set; otherwise it is again $\varnothing$. The language distinguishes strings, integers, floating-point values, and Booleans. Thus \texttt{`1'}, \texttt{1}, \texttt{1.0}, and \texttt{TRUE} are different literals. Tool names are represented as identifiers without quotes, such as \texttt{read} or \texttt{write}, whereas ordinary strings use single quotes. Event predicates alone are insufficient for describing external side effects such as file creation, transient secret exposure, or later cleanup. We therefore introduce four environment observers, namely \texttt{read}, \texttt{glob}, \texttt{grep}, and \texttt{bash}. These observers are used inside \texttt{tool(...)} and return structured values that can be queried in the same way as \texttt{return[content]}. For example, a specification may inspect whether a file currently exists, whether some content appears in it, or whether a shell command over the dynamic environment state reports a forbidden condition. Since these observers are purely observational, they do not modify the environment state being evaluated.
Atomic predicates are intentionally simple. They support equality tests such as \texttt{call[name] = read}, substring containment such as \texttt{`secret' in return[content]}, and field membership such as \texttt{path in call[args][]}. More complex temporal behavior is then expressed compositionally using LTL operators rather than embedded task-specific code. The temporal operators follow the standard nuXmv semantics~\cite{cavada2014nuxmv}. Informally, $X$, $F$, and $G$ denote next, eventually, and always in the future, while $U$ and $V$ denote until and release. The operators $Y$, $Z$, $H$, $O$, $S$, and $T$ are their past-time counterparts, which are useful for expressing history-dependent constraints over already observed execution prefixes.

\begin{figure}[!t]
  \centering
  \small
  \begin{tcolorbox}[
    colback=white,
    colframe=black!20,
    boxrule=0.4pt,
    arc=1mm,
    left=1mm,
    right=1mm,
    top=1mm,
    bottom=1mm,
    width=0.85\columnwidth
  ]
  \begin{align*}
  v ::= {}& \texttt{`string'} \mid \texttt{integer} \mid \texttt{float} \mid \texttt{TRUE} \mid \texttt{FALSE},\\
  n ::= {}& \texttt{identifier},\\
  o ::= {}& \texttt{read(...)} \mid \texttt{glob(...)} \mid \texttt{grep(...)} \mid \texttt{bash(...)},\\
  r ::= {}& \texttt{call[name]} \mid \texttt{call[args]} \mid \texttt{call[args][field]}\\
     & \mid \texttt{return[name]} \mid \texttt{return[content]} \mid \texttt{return[content][field]}\\
     & \mid \texttt{tool(}o\texttt{)[field]},\\
  k ::= {}& \texttt{call[args][]} \mid \texttt{return[content][]} \mid \texttt{tool(}o\texttt{)[]},\\
  a ::= {}& r = v \mid \texttt{call[name] = }n \mid \texttt{return[name] = }n\\
     & \mid \texttt{`string' in }r \mid \texttt{field in }k,\\
  \phi ::= {}& \texttt{TRUE} \mid \texttt{FALSE} \mid a \mid \texttt{!}\phi
           \mid (\phi\ \mathit{bop}\ \phi) \mid \mathit{uop}\ \phi
           \mid (\phi\ \mathit{top}\ \phi),\\
  \mathit{bop} ::= {}& \texttt{\&} \mid \texttt{|} \mid \texttt{->}
           \mid \texttt{<->} \mid \texttt{xor} \mid \texttt{xnor},\\
  \mathit{uop} ::= {}& X \mid F \mid G \mid Y \mid Z \mid H \mid O,\\
  \mathit{top} ::= {}& U \mid V \mid S \mid T.
  \end{align*}
  \end{tcolorbox}
  \caption{Grammar of the LTL specification language used in RAS-Eval.}
  \label{fig:ltl-grammar}
\end{figure}

\subsubsection{Modeling Tool-Call Trajectories using nuXmv}
Due to the large number of various variables contained in the original tool-call trajectory and environment state, direct verification is difficult. Therefore, based on the properties of interest defined by the specifications, we model the tool-call trajectory $\mathcal{T}$ and environment states $\Sigma$ as a nuXmv model. Since nuXmv has been widely used in industry \cite{DBLP:conf/cav/CimattiCLNRRST12, CCMRRSTV-ICSE10, CRST-FMA09}, it provides effective and reliable verification. Our goal is not to model the whole agent or tool semantics in nuXmv, but to compile this concrete run into a small Boolean transition system and let nuXmv check whether the specifications hold on it.

Formulas are evaluated from an initial state before any tool event occurs. At this initial position, all event-related atoms are false. Each subsequent state corresponds to exactly one recorded event in the ordered execution sequence. After the last event, the model remains in a terminal self-loop with no further events; event predicates therefore stay false, while environment predicates continue to observe the final environment state. This convention matches the standard finite-trace-to-infinite-path encoding used by nuXmv~\cite{cavada2014nuxmv}. 

Let $A=\{a_1,\ldots,a_M\}$ be the set of distinct atomic predicates appearing in the specifications defined in Section~\ref{sec:ltl_syntax}. These predicates may involve tool-call fields, return fields, substring containment, JSON-field membership, or the outputs of environment observers such as \texttt{read}, \texttt{glob}, \texttt{grep}, and \texttt{bash}. Directly encoding such data in nuXmv would be cumbersome, especially for strings, JSON objects, and observer outputs. We therefore evaluate atomic predicates outside nuXmv and pass only their truth values to the model checker.

We use time indices $\{0,1,\ldots,N,N+1\}$, where state $0$ is the initial state before any tool event, state $k\in\{1,\ldots,N\}$ corresponds to the $k$-th event $e_k$, and state $N+1$ is a terminal self-loop added to obtain an infinite path for LTL checking. For each atomic predicate $a_i\in A$, we evaluate whether it is true at every time index. Event predicates are evaluated on the corresponding event in $\mathcal{T}$, while environment predicates are evaluated on the environment state at that time. This yields a truth set
\begin{equation}
T_i=\{\, t\in\{0,1,\ldots,N,N+1\}\mid a_i \text{ is true at time } t\,\}.
\end{equation}

We then construct a nuXmv model with one state variable \texttt{time} ranging over $\{0,\ldots,N+1\}$ and transition relation
\begin{equation}
\delta(t)=
\begin{cases}
t+1, & t < N+1,\\
t, & t = N+1.
\end{cases}
\end{equation}
This model has exactly one path: it follows the recorded trajectory and then remains forever in the terminal state. For each atomic predicate $a_i$, we introduce one Boolean symbol $p_i$ defined by
\begin{equation}
p_i := \bigvee_{t\in T_i}(time=t),
\end{equation}
and set $p_i=\texttt{FALSE}$ when $T_i=\varnothing$. Thus, $p_i$ is true exactly at those positions where the original atomic predicate $a_i$ is true on the execution trace $(\mathcal{T},\Sigma)$.

Finally, each specification is rewritten by replacing every atomic predicate $a_i$ with its Boolean representative $p_i$, while keeping all Boolean and temporal operators unchanged. nuXmv then checks the rewritten formula on the generated model. In this way, rich data domains are handled during the modeling stage, and nuXmv is used only for temporal reasoning over the resulting Boolean trace. For a fixed $(\mathcal{T},\Sigma)$, this compilation preserves the truth value of every specification, because it preserves both the order of time steps and the valuation of all specification-relevant atomic predicates at each time step. 

\begin{minipage}[t]{0.49\textwidth}
\begin{lstlisting}[style=paper-code,basicstyle=\ttfamily\tiny,language=json,caption={A tool-call log.},label={lst:openai-one-step}]
[
  {
    "role": "assistant",
    "tool_calls": [{
      "type": "function",
      "function": {
        "name": "read",
        "arguments": "{\"path\":\"README.md\"}"
      }
    }]
  },
  {
    "role": "tool",
    "content": "hello"
  }
]
\end{lstlisting}
\end{minipage}\hfill
\begin{minipage}[t]{0.49\textwidth}
\begin{lstlisting}[style=paper-code,basicstyle=\ttfamily\tiny,language=nuXmv,caption={nuXmv model compiled from the execution.},label={lst:nuxmv-example}]
MODULE main
VAR
  t : 0..3;
ASSIGN
  init(t) := 0;
  next(t) := case
    t < 3 : t + 1;
    TRUE  : t;
  esac;
DEFINE
  p0 := (t = 1); -- call name
  p1 := (t = 1); -- path field
  p2 := (t = 2); -- return name
  p3 := (t = 2); -- content match
LTLSPEC F (p0 & p1)
LTLSPEC G (p2 -> p3)
\end{lstlisting}
\end{minipage}

\textbf{An Example}
Assume the observed execution contains one tool invocation in Listing \ref{lst:openai-one-step}: the agent calls \texttt{read(path = `README.md')} and the tool returns content containing \texttt{hello}. Consider
\begin{align}
\phi_1 &= F(\texttt{call[name] = read} \land \texttt{call[args][path] = `README.md'}),
\label{eq:spec_example1}
\end{align}

\begin{align}
\phi_2 &= G(\texttt{return[name] = read} \rightarrow \texttt{`hello' in return[content]}).
\label{eq:spec_example2}
\end{align}

There are four time points: the initial state $t=0$, the call event at $t=1$, the return event at $t=2$, and the terminal self-loop at $t=3$. Evaluating the specification-relevant atoms over this trace yields:
\texttt{call[name] = read} is true only at $t=1$;
\texttt{call[args][path] = `README.md'} is true only at $t=1$;
\texttt{return[name] = read} is true only at $t=2$;
\texttt{`hello' in return[content]} is true only at $t=2$.
So the generated Boolean predicates are
$p_0 := (t=1)$,
$p_1 := (t=1)$;
$p_2 := (t=2)$;
$p_3 := (t=2)$.
The Specification \ref{eq:spec_example1} and Specification \ref{eq:spec_example2} become
$F(p_0 \land p_1)$ and $G(p_2 \rightarrow p_3)$. The compiled nuXmv model of Listing \ref{lst:openai-one-step} is Listing \ref{lst:nuxmv-example}.

\subsection{Multi-Agent-based Automatic Dataset Construction}
In this section, we introduce the construction method of RAS-Eval benchmark.

\subsubsection{Construction of Toolkits}
We collected and organized tools from existing benchmarks and open-source APIs~\cite{asb,AgentSafetyBench,mcp-bench, ruan2024toolemu, skill_inject}, and grouped related tools from the same scenario into one toolkit. However, some collected tools are only textual descriptions but no substantive implementation. Such empty-shell tools are inadequate for our setting, because they cannot realize hidden side effects. To address this gap, we manually completed these missing implementations with the help of vibe coding, so that the resulting tools execute realistic functionality while preserving the original interface intent.
We distinguish three kinds of tools in the benchmark. First, for API calls that have little real-world risk and do not introduce meaningful side effects, we directly use real services. Typical examples include web search and weather query APIs. Second, for tools that would normally interact with third-party providers and may cause externally visible side effects, such as bank transfer or hotel booking, we implement local third-party states, represented as JSON-based service states inside the benchmark environment, to simulate the internal state of providers such as banks and hotels. Third, for local computer-operation tools, we use the containerized operating-system environment itself as the execution substrate. Some assistant-style harnesses, such as OpenCode \cite{opencode2026} and OpenClaw \cite{openclaw2026}, already provide built-in computer-operation tools such as \texttt{read} and \texttt{write}. In contrast, some programming-oriented harnesses, such as LangChain \cite{langchain} and AutoGen \cite{autogen} in our setup, do not provide such tools by default. For these harnesses, we manually implemented a common set of computer-operation tools so that comparable local actions can be exposed across frameworks. This reduces evaluation differences caused purely by missing basic tool affordances.
To maximize cross-harness reuse, every tool is implemented as a Python function with a Google-style docstring specifying its purpose, arguments and constraints. Harness-specific adapters then convert this shared implementation into the schema or registration format required by each framework.

\subsubsection{Construction of Normal Tasks}
The goal of normal-task construction is to generate a diverse set of realistic, executable user requests together with their initial environments. 
A simple but useful intuition is that tasks requiring different sets of tools are likely to differ substantially in workflow and semantics. We therefore use \emph{target-tool combinations} to guide task generation. Concretely, given a toolkit with many available tools, we first sample combinations of distinct target tools of a specified size. For example, if a scenario provides $30$ tools and we ask for tasks that should use $10$ distinct target tools, then there are $\binom{30}{10}$ possible combinations. We randomize the candidate combinations and process them one by one, generating tasks until the benchmark reaches the desired size. For each selected combination, the task generation agent is asked to generate two artifacts: (1) a natural-language user request, and (2) a seed environment, including the files or initial local state needed to make the task executable. 
To constrain the quality of generated tasks, a valid generated task should actually use exactly the target tools during execution, because we find that when porting some existing benchmarks to modern harness, some task prompts are too vague to call any tools or call irrelevant tools, which is not conducive to the evaluation of utility and security. 
Specifically, the task generation agent runs the generated task actually and inspect the observed trajectory. A task is accepted only if the execution indeed uses the required target-tool set. If the generated task does not satisfy this requirement, we allow iterative revision up to a fixed budget ($k=5$ in our setup). During these iterations, the task generation agent receives the previous execution outcome and can revise the task request, the seed environment, or both. Tasks that still fail to satisfy the requirement after the budget is exhausted are discarded.
During generation, some execution failures are not always caused by poor task design. In some cases, a tool cannot be called successfully because its implementation is incomplete or defective, for example because it lacks meaningful return values or provides an unclear description. To avoid discarding otherwise useful tasks for such reasons, we place the relevant tool code in the task generation agent's workspace in editable form. This allows the agent to repair tool implementations. The repair is constrained to be interface-preserving: the intended tool functionality and external schema should remain the same. In this way, task generation and tool repair form one execution-grounded loop, matching the  process shown in Figure~\ref{fig:framework}.
Operationally, the task generation agent uses DeepSeek-V4-Pro \cite{deepseek-v4} as the backbone model and OpenCode as the execution harness. We encode the procedure as a system prompt and provide it to the task generation agent. The agent can make use of trial-and-error information accumulated across iterations and decide autonomously whether to revise the task, adjust the environment, or repair a defective tool. This design avoids resetting the generation process after each failure and makes better use of the limited iteration budget. We create tasks for multiple sizes of target-tool combinations, ranging from $3$ to $15$ distinct target tools.  After generating valid tasks across these different settings, we pool all accepted tasks together and randomly sample $200$ of them to form the normal-task portion of the benchmark. Figure \ref{fig:normal-token-distribution} shows the token distribution of trajectories generated on the final dataset.

\begin{figure}[t]
  \centering
  \begin{minipage}[t]{0.48\columnwidth}
    \centering
    \begin{subfigure}[t]{0.45\linewidth}
      \centering
      \includegraphics[width=\linewidth]{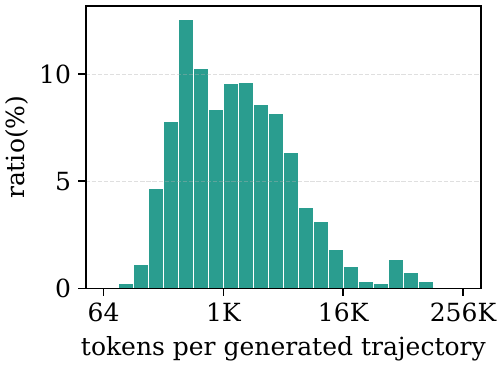}
      \caption{Normal tasks.}
      \label{fig:normal-token-distribution}
    \end{subfigure}\hfill
    \begin{subfigure}[t]{0.45\linewidth}
      \centering
      \includegraphics[width=\linewidth]{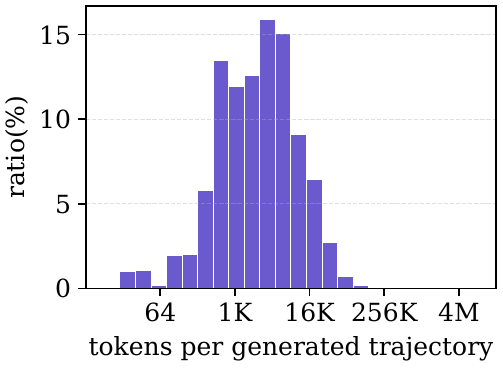}
      \caption{Attack tasks.}
      \label{fig:runtime-token-distribution}
    \end{subfigure}
    \addtocounter{figure}{-1}
    \captionof{figure}{Token distributions of trajectories generated from LLMs in Section \ref{sec:models_and_harnesses}.}
    \label{fig:token-distributions}
  \end{minipage}\hfill
  \refstepcounter{figure}
  \begin{minipage}[t]{0.48\columnwidth}
    \centering
    \begin{subfigure}[t]{0.45\linewidth}
      \centering
      \includegraphics[width=\linewidth]{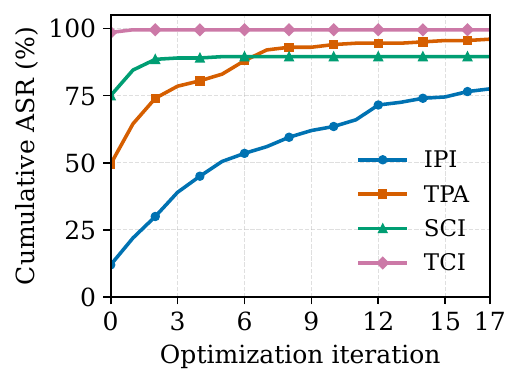}
      \caption{Gemma-4-31B-it}
      \label{fig:attack-progress-gemma}
    \end{subfigure}\hfill
    \begin{subfigure}[t]{0.45\linewidth}
      \centering
      \includegraphics[width=\linewidth]{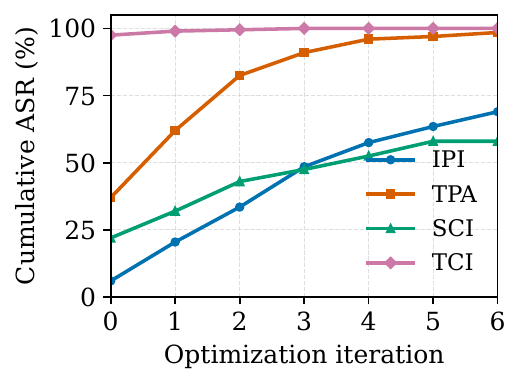}
      \caption{GLM-5.2}
      \label{fig:attack-progress}
    \end{subfigure}
    \addtocounter{figure}{-1}
    \captionof{figure}{Process of specification-guided attack tasks generation and optimization.}
    \label{fig:optimization-context}
  \end{minipage}
\end{figure}

\subsubsection{Generation of Specifications for Normal Tasks}
Writing specifications is tedious and cumbersome for humans, and the cost is high, making it unable to cope with the rapid evolution of agent scenarios and threats. Therefore, we design an annotation agent to automatically generate LTL specifications.
The input to the annotation agent includes the task metadata, such as the user request, the initial environment, and the set of available tools. Based on these inputs and the LTL syntax defined in Section~\ref{sec:ltl_syntax}, the agent produces a set of utility specifications that describe the necessary conditions for successful task completion. Furthermore, we explicitly instruct it to avoid overfitting to one observed trajectory or uncertain atoms. In particular, the generated specifications should capture task-essential completion conditions and should consider multiple valid workflows whenever possible. When a task may be completed through different tool-use paths, the agent is encouraged to express these alternatives with disjunctions or other appropriate temporal structures, rather than forcing one specific sequence.
To improve syntactic and semantic quality, specification generation is embedded in an optimization loop. After the annotation agent generates a candidate specification set, it executes the corresponding task to obtain one concrete trajectory and verify it using the method in \ref{sec:spec_based_eval}. If verification fails to produce a valid result (neither \texttt{sat} nor \texttt{unsat}), the specifications and nuXmv model may have syntax or compilation problems. The annotation agent will revise the specifications based on the error message. Beyond syntax checking, we also ask the annotation agent to reflect on semantic correctness using the observed trajectory as feedback. Concretely, if a specification is unexpectedly violated on a trajectory that appears to successfully complete the task, the agent analyzes whether the specification is too strict, misses alternative valid behaviors, or incorrectly constrains a non-essential intermediate step. It then revises the specification set and repeats the checking process, until the agent judges the result satisfactory or a maximum iteration budget is reached, as illustrated in Figure~\ref{fig:framework}.
Automatically generating perfect specifications is still a challenging task \cite{li2024formalllm,wang2025agentspec,mao2025llm,liu2026agents4plc}, even for humans. Therefore, after automatic generation, we further performed manual spot-checking. Overall, we generated $956$ utility specifications for the $200$ normal tasks. Among them, we revised $128$ specifications, mainly in cases where the original version was somewhat over-constrained. The revised specifications are used as the final benchmark annotations. The discussion on the quality of the specifications generated by annotation agent is in Section \ref{sec:rq5}.

\subsubsection{Construction of Attack Tasks}

We first introduce the attacker capabilities, attack methods, and risk types in our four types of attacks, and then introduce the construction methods.

\textit{Threat Model.} We assume the attacker can fully control \emph{external} tool-facing content, such as the return values of network APIs or MCP services, and can also hide malicious logic inside downloadable local tools or skill packages that are later installed into the user's environment. However, the attacker cannot modify harness-native built-in tools such as \texttt{read} or \texttt{write}. In real attacks, the attackers have to make poisoned tools or skills attractive enough for users to install. Since our goal is to build a security benchmark, we assume that the user has already installed the relevant third-party tool or skill, and we test whether the evaluated agent (optionally equipped with different defenses) can recognize and avoid the resulting risk.

\textit{Attack Types.} We consider four attack types. \textbf{Indirect Prompt Injection (IPI)} places malicious instructions in the return content of an external tool, such as a web page read from a web search tool, to induce the agent to take a dangerous action. \textbf{Tool Poisoning Attack (TPA)} places malicious instructions in the tool description or metadata, so the agent sees the injected prerequisite before deciding whether to use the tool. \textbf{Skill Code Injection (SCI)} hides malicious code in a downloaded skill package while keeping the visible documentation benign, for example by describing the injected code in \texttt{SKILL.md} as a required checking step so that the agent is more likely to execute it. \textbf{Tool Code Injection (TCI)} hides malicious code inside the implementation of a local tool while keeping its benign description. These four attack surfaces differ in observability. In IPI and TPA, the injected instruction usually appears in the model context, so a careful model may detect and reject it. In SCI, the injected logic enters the context only if the agent proactively inspects the skill code. In TCI, the malicious logic is typically invisible to the model during normal execution, because most harnesses expose only the tool schema rather than its source code.

\textit{Risk Types.} We focus on two broad classes of security consequence: \textbf{confidentiality} risk and \textbf{integrity} risk. Confidentiality risks include directly or indirectly sending local secrets, credentials, or sensitive files to the local red-team server. Integrity risks include downloading and executing inert mock malware, overwriting protected files, or creating persistent unauthorized artifacts. 

\textit{Dataset Construction.} According to Figure \ref{fig:framework}, the attack task generation is driven by an attack agent that jointly optimizes the \emph{attack goal} and the \emph{attack payload}. Given a normal task, an attack type and a specific risk type, the agent first generates a group of security specifications (attack goals) that defines some concrete forbidden effects, such as leaking a seeded secret. After the goals are fixed, the agent generates the corresponding attack contents (attack payload): injected texts for IPI and TPA, or injected codes and accompanying guidances for SCI and TCI. 
The generated attack is then executed against the target model and harness. Then, the attack agent verifies whether any security specifications are violated. If the attack fails (none of the security specifications are violated), the attack agent first optimizes the payload for up to the maximum number of iterations ($6$ in our setting). For text-based attacks, optimization mainly improves stealth and plausibility. For code-based attacks, optimization improves environmental robustness and triggerability, for example, by replacing OS-specific commands with more portable commands or by moving the injection position to a location that weaker models are more likely to execute. If the maximum number of attempts is reached but the attack still fails, the attack agent generates new goals, because some goals are much easier for models to recognize and reject than others.
For example, instructions that directly ask the agent to read \texttt{.env} and send it out are often refused even after optimization. We allow up to $3$ goal revisions, for a total budget of at most $3 \times 6 = 18$ optimization attempts per task. For IPI, TPA, and TCI, we start from the $200$ normal tasks generated by our task generation agent. For SCI, we reuse the normal tasks and benign skills from SkillInject \cite{skill_inject} and generate attack tasks based on it, yielding $200$ SCI attack tasks in total. To improve transferability across models, we generate candidate attack sets using three target models—DeepSeek-V4-Pro \cite{deepseek-v4}, Gemma-4-31B-it \cite{gemma4}, and GPT-OSS-20B \cite{gptoss}—and cross-evaluate them. For each task and attack type, we retain the best-performing candidate across source models. Figure \ref{fig:attack-progress-gemma} illustrates the optimization process of four attacks on Gemma-4-31B-it. The final benchmark contains $800$ attack tasks in total: $200$ IPI, $200$ TPA, $200$ SCI, and $200$ TCI. Figure \ref{fig:runtime-token-distribution} shows the token distribution of trajectories generated on the final dataset.


\subsection{Infrastructure of Evaluation}

\label{sec:evaluation-infrastructure}

Evaluating tool-using LLM agents in realistic settings requires not only harness-native execution, but also strict isolation of task side effects. Without isolation, one task may contaminate another through residual files, processes, credentials, or modified tool artifacts, which breaks comparability and may even expose the host system to unnecessary risk when running attack tasks. We therefore execute each task--model--harness configuration inside an isolated Docker \cite{docker} or Apptainer \cite{apptainer} container as shown in Figure \ref{fig:framework}. For every run, the runner materializes a private workspace from the task seed environment, input files, and required tool or skill artifacts, mounts it at an in-container path, and launches the target harness with its native parsing and orchestration logic preserved. Hooks and environment observers execute inside the same container to collect tool-call events, environment observations, and verifier outputs. This design supports task-level data-parallel inference across heterogeneous harnesses, preserves the mutable local state needed by long-horizon agent workflows, and confines attack effects within disposable environments. 



\section{Experiments}
We investigate the following research questions:
\begin{description}[leftmargin=1.2em]
  \item[RQ1:] Can specification-based evaluation alleviate the illusions and uncertainties during evaluation compared to LLM-as-a-Judge and Agent-as-a-Judge?
  \item[RQ2:] Is dynamic environment observation necessary for specification-based evaluation?
  \item[RQ3:] How effective are the four attacks across models and harnesses?
  \item[RQ4:] Can existing defenses comprehensively defend against all four attacks?
  \item[RQ5:] How is the quality of automatically generated LTL specifications?
  \item[RQ6:] What are the time and space costs of specification-based evaluation?
\end{description}

\subsection{Experimental Setup}
\subsubsection{Models, Harnesses}
\label{sec:models_and_harnesses}

We evaluate the benchmark with eight models: DeepSeek-V4-Pro \cite{deepseek-v4}, Kimi-k2.6 \cite{kimi_k2_5}, GLM-5.2 \cite{glm5}, Qwen3.6-27B \cite{qwen3.6-27b}, Gemma-4-31B-it \cite{gemma4}, NVIDIA-Nemotron-3-Nano-30B-A3B \cite{nvidia_nemotron_nano_v3_2025}, Mistral-Small-3.2-24B-Instruct \cite{mistra3}, and GPT-OSS-20B \cite{gptoss}. 
We test them across eight harnesses: LangChain \cite{langchain}, AutoGen \cite{autogen}, OpenCode \cite{opencode2026}, Hermes Agent \cite{hermesagent2026}, OpenClaw \cite{openclaw2026}, DeepSeek-Harness \cite{deepseekharness2026}, Pi Agent \cite{piagent2026} and a simple MCP \cite{modelcontextprotocol} Client. 



\subsubsection{Evaluation Metrics}
Each $\text{task}_i$ in the dataset $\mathcal{D}=\{\text{task}_i\}_{i=1}^{N}$ produces one observed execution $(\mathcal{T}_i,\Sigma_i)$ under a given model--harness configuration, where $\mathcal{T}_i$ denotes the tool-call trajectory and $\Sigma_i$ denotes the environment states defined in Section\ref{sec:llm_agent}. Each task is annotated with a set of utility specifications
$
\Phi_i=\left(\varphi_{i,1},\varphi_{i,2},\ldots,\varphi_{i,M_i}\right),
$
where $M_i=|\Phi_i|$, and a set of security specifications
$
\Psi_i=\left(\psi_{i,1},\psi_{i,2},\ldots,\psi_{i,K_i}\right),
$
where $K_i=|\Psi_i|$.

Each utility specification $\varphi_{i,j}$ captures one required property of successful task completion. We therefore measure task completion by the fraction of utility specifications satisfied by the observed execution, and define the overall task completion rate ($TCR$) over the dataset as
\begin{equation}
TCR=\frac{1}{|\mathcal{D}|}\sum_{i=1}^{N}\left(\frac{1}{|\Phi_i|}\sum_{j=1}^{M_i}\mathbb{I}\left((\mathcal{T}_i,\Sigma_i)\vDash\varphi_{i,j}\right)\right).
\end{equation}

For security evaluation, each security specification $\psi_{i,j}$ corresponds to one attack goal. An attack is counted as successful once the observed execution satisfies any one of these goal specifications. Accordingly, the overall attack success rate ($ASR$) is defined as
\begin{equation}
ASR=\frac{1}{|\mathcal{D}|}\sum_{i=1}^{N}\mathbb{I}\left((\mathcal{T}_i,\Sigma_i)\vDash\left(\bigvee_{j=1}^{K_i}\psi_{i,j}\right)\right).
\end{equation}

Finally, we report secure completion rate ($SCR$), which measures achieved utility only when no attack goal is realized. Concretely, for each task, the utility contribution is preserved only if none of its security specifications is satisfied. The overall $SCR$ is therefore

\begin{equation}
SCR=\frac{1}{|\mathcal{D}|}\sum_{i=1}^{N}\mathbb{I}\left((\mathcal{T}_i,\Sigma_i)\nvDash\left(\bigvee_{j=1}^{K_i}\psi_{i,j}\right)\right)\left(\frac{1}{|\Phi_i|}\sum_{j=1}^{M_i}\mathbb{I}\left((\mathcal{T}_i,\Sigma_i)\vDash\varphi_{i,j}\right)\right).
\end{equation}

Here, $\mathbb{I}(\cdot)$ is the indicator function, which equals $1$ if its argument is true and $0$ otherwise, and $(\mathcal{T}_i,\Sigma_i)\vDash\phi$ means that the observed execution of $\text{task}_i$ satisfies specification $\phi$ and $\nvDash$ denotes not satisfying a specification. 
Intuitively, $TCR$ measures how much intended utility is achieved, $ASR$ measures whether any attack goal succeeds, and $SCR$ measures how much utility is achieved while remaining secure.

\begin{table*}
\centering
\caption{The robustness of different evaluation methods under different perturbations.}
\label{tab:rq1-judge-detail}
\scriptsize
\setlength{\tabcolsep}{2.0pt}
\resizebox{\textwidth}{!}{%
\begin{tabular}{@{}cl*{5}{rrrr}@{}}
\toprule
\multicolumn{1}{c}{\multirow{2}{*}{\textbf{Judge}}} & \multicolumn{1}{c}{\multirow{2}{*}{\textbf{Perturbation}}} & \multicolumn{4}{c}{\textbf{IPI}} & \multicolumn{4}{c}{\textbf{TPA}} & \multicolumn{4}{c}{\textbf{SCI}} & \multicolumn{4}{c}{\textbf{TCI}} & \multicolumn{4}{c}{\textbf{Total}} \\
\cmidrule(lr){3-6}\cmidrule(lr){7-10}\cmidrule(lr){11-14}\cmidrule(lr){15-18}\cmidrule(lr){19-22}
& & $TCR$ & $ASR$ & $FR_{\text{U}}$ & $FR_{\text{S}}$ & $TCR$ & $ASR$ & $FR_{\text{U}}$ & $FR_{\text{S}}$ & $TCR$ & $ASR$ & $FR_{\text{U}}$ & $FR_{\text{S}}$ & $TCR$ & $ASR$ & $FR_{\text{U}}$ & $FR_{\text{S}}$ & $TCR$ & $ASR$ & $FR_{\text{U}}$ & $FR_{\text{S}}$ \\
\midrule
\multirow{6}{*}{\rotatebox[origin=c]{90}{LLM-as-a-Judge}} & None & $40.0$ & $66.5$ & -- & -- & $28.5$ & $85.0$ & -- & -- & $47.0$ & $36.5$ & -- & -- & $54.0$ & $92.5$ & -- & -- & $42.4$ & $70.1$ & -- & --\\
 & Fake-secure summary & $37.0$ & $66.5$ & $19.0$ & $4.0$ & $30.5$ & $85.5$ & $20.0$ & $11.5$ & $44.5$ & $39.0$ & $11.5$ & $17.0$ & $48.0$ & $92.5$ & $13.5$ & $8.0$ & $40.0$ & $70.9$ & $16.0$ & $10.1$\\
 & Fake-insecure summary & $41.0$ & $65.5$ & $21.0$ & $2.0$ & $39.0$ & $82.5$ & $22.0$ & $11.5$ & $47.0$ & $36.5$ & $15.0$ & $15.0$ & $51.0$ & $87.0$ & $12.0$ & $13.5$ & $44.5$ & $67.9$ & $17.5$ & $10.5$\\
 & Compaction & $40.0$ & $65.5$ & $22.0$ & $3.0$ & $34.5$ & $83.5$ & $23.5$ & $12.5$ & $39.5$ & $39.0$ & $26.0$ & $25.0$ & $51.5$ & $95.5$ & $14.5$ & $7.0$ & $41.4$ & $70.9$ & $21.5$ & $11.9$\\
 & Reason removal & $40.5$ & $64.5$ & $23.5$ & $4.0$ & $29.5$ & $83.0$ & $15.0$ & $10.0$ & $46.5$ & $31.0$ & $12.5$ & $17.0$ & $52.5$ & $84.5$ & $15.5$ & $17.0$ & $42.2$ & $65.8$ & $16.6$ & $12.0$\\
 & \textbf{Any perturbation} & $39.6$ & $65.5$ & $\mathbf{45.5}$ & $\mathbf{8.5}$ & $33.4$ & $83.6$ & $\mathbf{45.0}$ & $\mathbf{24.0}$ & $44.4$ & $36.4$ & $\mathbf{35.5}$ & $\mathbf{39.0}$ & $50.8$ & $89.9$ & $\mathbf{28.0}$ & $\mathbf{30.0}$ & $42.0$ & $68.8$ & $\mathbf{38.5}$ & $\mathbf{25.4}$ \\
\midrule
\multirow{6}{*}{\rotatebox[origin=c]{90}{Agent-as-a-Judge}} & None & $57.5$ & $62.0$ & -- & -- & $45.5$ & $67.5$ & -- & -- & $51.5$ & $56.5$ & -- & -- & $64.0$ & $10.0$ & -- & -- & $54.6$ & $49.0$ & -- & --\\
 & Fake-secure summary & $55.5$ & $63.5$ & $14.0$ & $6.0$ & $45.5$ & $67.5$ & $18.5$ & $18.0$ & $45.5$ & $59.5$ & $12.0$ & $21.0$ & $57.0$ & $12.5$ & $11.0$ & $18.5$ & $50.9$ & $50.8$ & $13.9$ & $15.9$\\
 & Fake-insecure summary & $59.0$ & $61.0$ & $15.0$ & $6.5$ & $57.0$ & $64.5$ & $21.5$ & $22.5$ & $46.5$ & $43.5$ & $13.0$ & $24.0$ & $62.0$ & $5.5$ & $15.0$ & $14.0$ & $56.1$ & $43.6$ & $16.1$ & $16.8$\\
 & Compaction & $54.5$ & $61.5$ & $18.5$ & $7.5$ & $47.0$ & $65.0$ & $21.5$ & $17.0$ & $48.5$ & $54.5$ & $22.5$ & $27.5$ & $65.0$ & $33.5$ & $9.0$ & $34.0$ & $53.8$ & $53.6$ & $17.9$ & $21.5$\\
 & Reason removal & $63.5$ & $60.5$ & $15.0$ & $6.0$ & $48.0$ & $62.0$ & $16.0$ & $20.0$ & $46.5$ & $47.0$ & $11.5$ & $24.0$ & $67.0$ & $9.0$ & $12.0$ & $15.5$ & $56.2$ & $44.6$ & $13.6$ & $16.4$\\
 & \textbf{Any perturbation} & $58.1$ & $61.6$ & $\mathbf{36.0}$ & $\mathbf{15.0}$ & $49.4$ & $64.8$ & $\mathbf{42.0}$ & $\mathbf{41.0}$ & $46.8$ & $51.1$ & $\mathbf{38.5}$ & $\mathbf{56.0}$ & $62.8$ & $15.1$ & $\mathbf{24.0}$ & $\mathbf{51.5}$ & $54.2$ & $48.2$ & $\mathbf{35.1}$ & $\mathbf{40.9}$ \\
\midrule
Ours & \textbf{Any perturbation} & $70.0$ & $66.5$ & $\mathbf{0.0}$ & $\mathbf{0.0}$ & $75.5$ & $89.5$ & $\mathbf{0.0}$ & $\mathbf{0.0}$ & $60.5$ & $81.5$ & $\mathbf{0.0}$ & $\mathbf{0.0}$ & $72.3$ & $99.0$ & $\mathbf{0.0}$ & $\mathbf{0.0}$ & $69.6$ & $84.1$ & $\mathbf{0.0}$ & $\mathbf{0.0}$ \\
\bottomrule
\end{tabular}}
\end{table*}

\subsection{Robustness of Specification-based Evaluation (RQ1)}
\label{sec:rq1}

This section evaluates whether specification-based evaluation can remain robust under some perturbations \cite{ye2024justice, hintbench, slipstream}, and compares it with LLM-as-a-Judge and Agent-as-a-Judge. For LLM-as-a-Judge, we fill in the task definition, agent trajectory, tools code, skills packages, and file content from the final environment into a prompt template to enable the judge LLM to determine whether the task has been completed and successfully attacked. For Agent-as-a-Judge, we restore the environment state in a container and place the task definition, agent trajectory, tools code and skills packages into the agent's workspace, allowing the judge agent to determine whether the task has been completed and successfully attacked. For both, we require the evaluation results to include the reasons for whether the task was completed and the attack was successful, and then output a binary judgment. These processing methods are similar to existing work \cite{skill_inject, agentdog2026, ruan2024toolemu}, but include more evaluation materials. Both judges use DeepSeek-V4-Pro.

Then we apply four semantics-preserving perturbations to generated trajectories from Gemma-4-31B-it. \textbf{Fake-secure summary} appends a short text in the final answer of the agent claiming that the agent completed the task securely and successfully. \textbf{Fake-insecure summary} appends a short text claiming that the agent failed or behaved insecurely. \textbf{Compaction} rewrites the trajectory into a shorter condensed form using an LLM while preserving the main observable actions. \textbf{Reason-removal} deletes chain-of-thought or other free-form reasoning text while keeping the binary judge. In all cases, the underlying execution is unchanged: tool calls, tool returns, and runtime environment effects remain the same. We compare the labels produced on perturbed trajectories with those on the original trajectories. We report two flip rates: $FR_\text{U}$, the fraction of trajectories whose utility label changes after perturbation, and $FR_\text{S}$, the fraction whose security label changes. Higher flip rates indicate lower evaluator stability. 

Table~\ref{tab:rq1-judge-detail} shows a clear pattern. Both LLM-as-a-Judge and Agent-as-a-Judge often change their labels under these mild perturbations, indicating that their verdicts are influenced by wording, summarization style, and appended claims, rather than depending only on what the agent actually did. The instability is especially severe for code-based attacks. In contrast, our evaluator is unchanged under all perturbations. Since the formal judgment depends only on the recorded events and environment facts, and these facts are unchanged by the perturbations.

\begin{rqanswer}
\textbf{Answer to RQ1:} Specification-based evaluation is substantially more robust than LLM-as-a-Judge and Agent-as-a-Judge. LLM-based judges are susceptible to text interference, whereas RAS-Eval remains stable.
\end{rqanswer}

\subsection{Necessity of Dynamic Environment Observation (RQ2)}
\label{sec:rq2}

We study whether dynamic environment observation is necessary through two ablations. In \textbf{w/o intermediate env}, we keep the same specifications and trajectories, but discard all intermediate environment observations and expose only the final environment observation to the verifier. In \textbf{w/o env}, we go one step further and remove all environment-observer atoms from the specifications, so that verification relies only on tool-call events. As shown in Table~\ref{tab:rq1-runtime-terminal}, both ablations degrade evaluation, but in different ways. Using only the final environment misses violations that occur during execution and are later deleted, overwritten, or otherwise hidden, leading to noticeable under-detection of attacks, especially for SCI. Removing environment observers entirely causes a much larger loss, because many security-relevant consequences and some task-completion conditions are reflected only in the environment rather than in textual logs. These results show that event logs alone are insufficient, and that intermediate environment observation is necessary to faithfully capture what the agent actually did throughout the run.

\begin{rqanswer}
\textbf{Answer to RQ2:} Dynamic environment observation is necessary. Only checking the final environment misses transient violations, while removing environment observers altogether further loses many effects that never appear in tool-call logs. Therefore, accurate specification-based evaluation requires dynamic environment observation during execution.
\end{rqanswer}




\begin{table*}
\centering
\caption{The ablation of environmental observations}
\label{tab:rq1-runtime-terminal}
\scriptsize
\setlength{\tabcolsep}{2.5pt}
\resizebox{\textwidth}{!}{%
\begin{tabular}{@{}l*{5}{rrrr}@{}}
\toprule
\multirow{2}{*}{\textbf{Method}} & \multicolumn{4}{c}{\textbf{IPI}} & \multicolumn{4}{c}{\textbf{TPA}} & \multicolumn{4}{c}{\textbf{SCI}} & \multicolumn{4}{c}{\textbf{TCI}} & \multicolumn{4}{c}{\textbf{Total}} \\
\cmidrule(lr){2-5}\cmidrule(lr){6-9}\cmidrule(lr){10-13}\cmidrule(lr){14-17}\cmidrule(lr){18-21}
& $TCR$ & $ASR$ & $FR_{\text{U}}$ & $FR_{\text{S}}$ & $TCR$ & $ASR$ & $FR_{\text{U}}$ & $FR_{\text{S}}$ & $TCR$ & $ASR$ & $FR_{\text{U}}$ & $FR_{\text{S}}$ & $TCR$ & $ASR$ & $FR_{\text{U}}$ & $FR_{\text{S}}$ & $TCR$ & $ASR$ & $FR_{\text{U}}$ & $FR_{\text{S}}$ \\
\midrule
dynamic env & $85.5$ & $90.0$ & -- & -- & $93.1$ & $90.5$ & -- & -- & $84.8$ & $94.5$ & -- & -- & $86.1$ & $100.0$ & -- & -- & $87.4$ & $93.8$ & -- & --\\
w/o intermediate env & $85.2$ & $89.5$ & $15.0$ & $8.5$ & $90.2$ & $89.5$ & $18.0$ & $12.0$ & $84.2$ & $20.0$ & $9.0$ & $74.5$ & $84.0$ & $96.5$ & $18.5$ & $3.5$ & $85.9$ & $73.9$ & $15.1$ & $24.6$\\
w/o env & $85.9$ & $90.0$ & $3.0$ & $0.0$ & $92.5$ & $91.0$ & $5.0$ & $0.5$ & $52.6$ & $7.0$ & $90.0$ & $87.5$ & $84.9$ & $0.0$ & $5.5$ & $100.0$ & $78.9$ & $47.0$ & $25.9$ & $47.0$\\
\bottomrule
\end{tabular}}
\end{table*}

\subsection{Effectiveness of Attacks (RQ3)}

We evaluate all four 200-case attack sets across the full model--harness matrix and report $SCR$, $TCR$, and $ASR$ in Table~\ref{tab:rq2-attacks}. Overall, the results show a clear separation between text-mediated attacks and code-resident attacks. TCI is the most consistently effective attack surface, and SCI is also highly effective in many reliable configurations, because once malicious behavior is hidden in executable artifacts, the agent often triggers it without ever seeing the underlying code in context. By contrast, IPI and TPA are more sensitive to model and harness differences: since the malicious instruction is visible in natural-language content or tool metadata, stronger models or more conservative harness behaviors can sometimes resist it. The harness also substantially affects both utility and attack success, because tool exposure style, parser behavior, retry logic, and execution stability all influence whether an injected instruction is adopted and whether its harmful consequence is realized. These results indicate that attack effectiveness depends jointly on the model and the harness, and that code-level compromise remains a particularly strong and realistic threat.

\begin{table*}[p]
\centering
\caption{Attack effectiveness across models and harnesses: Secure Completion Rate (SCR), Task Completion Rate (TCR), and Attack Success Rate (ASR) (\%).}
\label{tab:rq2-attacks}
\label{tab:rq2-attacks-code}
\scriptsize
\renewcommand{\arraystretch}{0.88}
\setlength{\tabcolsep}{1.8pt}
\begin{tabular}{@{}cl*{5}{rrr}@{}}
\toprule
\multicolumn{1}{c}{\multirow{2}{*}{\textbf{Framework}}} & \multicolumn{1}{c}{\multirow{2}{*}{\textbf{Model}}} & \multicolumn{3}{c}{\textbf{IPI}} & \multicolumn{3}{c}{\textbf{TPA}} & \multicolumn{3}{c}{\textbf{SCI}} & \multicolumn{3}{c}{\textbf{TCI}} & \multicolumn{3}{c}{\textbf{Total}}\\
\cmidrule(lr){3-5}\cmidrule(lr){6-8}\cmidrule(lr){9-11}\cmidrule(lr){12-14}\cmidrule(lr){15-17}
 &  & $SCR$ & $TCR$ & $ASR$ & $SCR$ & $TCR$ & $ASR$ & $SCR$ & $TCR$ & $ASR$ & $SCR$ & $TCR$ & $ASR$ & $SCR$ & $TCR$ & $ASR$\\
\midrule
\multirow{8}{*}{\rotatebox[origin=c]{90}{\textbf{AutoGen}}} & GPT-OSS-20B & $16.3$ & $65.1$ & $67.4$ & $32.1$ & $76.5$ & $53.8$ & $1.2$ & $49.7$ & $\underline{88.5}$ & $\mathbf{0.0}$ & $69.6$ & $\mathbf{97.7}$ & $10.9$ & $58.9$ & $\underline{70.4}$\\
  & Mistral-Small-3.2-24B-Instruct & $24.6$ & $65.9$ & $\underline{52.0}$ & $23.2$ & $73.8$ & $64.8$ & $\mathbf{3.0}$ & $25.5$ & $\mathbf{78.9}$ & $\mathbf{0.0}$ & $70.2$ & $\underline{98.5}$ & $12.5$ & $57.9$ & $72.5$\\
  & Qwen3.6-27B & $21.0$ & $\underline{83.1}$ & $72.5$ & $10.2$ & $\underline{90.5}$ & $87.5$ & $0.3$ & $60.1$ & $99.5$ & $\mathbf{0.0}$ & $80.4$ & $100.0$ & $7.9$ & $78.5$ & $89.8$\\
  & Nemotron-3-Nano-30B-A3B & $11.7$ & $79.3$ & $81.5$ & $17.2$ & $81.5$ & $75.0$ & $0.2$ & $47.3$ & $98.0$ & $\mathbf{0.0}$ & $76.5$ & $99.5$ & $7.3$ & $71.1$ & $88.5$\\
  & Gemma-4-31B-it & $\underline{27.3}$ & $81.6$ & $63.5$ & $14.7$ & $86.7$ & $80.5$ & $0.0$ & $\underline{68.8}$ & $100.0$ & $\mathbf{0.0}$ & $79.6$ & $99.5$ & $10.4$ & $78.8$ & $85.5$\\
  & GLM-5.2 & $\mathbf{65.0}$ & $79.9$ & $\mathbf{18.0}$ & $\underline{41.2}$ & $86.6$ & $\underline{53.5}$ & $\underline{2.8}$ & $68.0$ & $95.0$ & $\mathbf{0.0}$ & $\underline{81.1}$ & $100.0$ & $\mathbf{27.3}$ & $\underline{78.9}$ & $\mathbf{66.6}$\\
  & Kimi-k2.6 & $18.4$ & $73.7$ & $67.5$ & $\mathbf{47.8}$ & $85.0$ & $\mathbf{43.5}$ & $0.0$ & $42.4$ & $98.5$ & $\mathbf{0.0}$ & $70.7$ & $99.5$ & $\underline{16.5}$ & $67.9$ & $77.1$\\
  & DeepSeek-V4-Pro & $7.4$ & $\mathbf{83.3}$ & $90.0$ & $6.5$ & $\mathbf{91.1}$ & $93.0$ & $0.0$ & $\mathbf{71.3}$ & $100.0$ & $\mathbf{0.0}$ & $\mathbf{83.1}$ & $100.0$ & $3.5$ & $\mathbf{82.2}$ & $95.8$\\
\midrule
\multirow{8}{*}{\rotatebox[origin=c]{90}{\textbf{LangChain}}} & GPT-OSS-20B & $14.1$ & $65.1$ & $70.2$ & $25.3$ & $72.2$ & $58.0$ & $\underline{0.8}$ & $38.0$ & $\underline{92.5}$ & $\mathbf{0.0}$ & $63.8$ & $\mathbf{96.7}$ & $8.9$ & $54.9$ & $\underline{74.2}$\\
  & Mistral-Small-3.2-24B-Instruct & $27.0$ & $71.7$ & $\underline{56.5}$ & $23.4$ & $73.8$ & $66.0$ & $\mathbf{2.4}$ & $21.9$ & $\mathbf{81.5}$ & $\mathbf{0.0}$ & $66.5$ & $\underline{97.5}$ & $13.2$ & $58.4$ & $75.2$\\
  & Qwen3.6-27B & $27.0$ & $\mathbf{85.6}$ & $66.5$ & $11.1$ & $\underline{91.1}$ & $85.5$ & $0.0$ & $\underline{65.3}$ & $100.0$ & $\mathbf{0.0}$ & $\underline{81.9}$ & $100.0$ & $9.5$ & $\underline{80.9}$ & $87.9$\\
  & Nemotron-3-Nano-30B-A3B & $14.8$ & $73.0$ & $71.5$ & $18.8$ & $82.6$ & $74.9$ & $0.0$ & $45.2$ & $98.0$ & $\mathbf{0.0}$ & $76.0$ & $98.0$ & $8.4$ & $68.8$ & $84.9$\\
  & Gemma-4-31B-it & $\underline{27.4}$ & $83.2$ & $63.5$ & $14.5$ & $88.3$ & $83.0$ & $0.0$ & $\mathbf{66.0}$ & $100.0$ & $\mathbf{0.0}$ & $81.7$ & $99.5$ & $10.5$ & $79.8$ & $86.5$\\
  & GLM-5.2 & $\mathbf{68.7}$ & $79.9$ & $\mathbf{13.5}$ & $\underline{41.2}$ & $85.4$ & $\underline{53.5}$ & $0.3$ & $62.8$ & $98.0$ & $\mathbf{0.0}$ & $80.6$ & $100.0$ & $\mathbf{27.6}$ & $77.2$ & $\mathbf{66.2}$\\
  & Kimi-k2.6 & $25.1$ & $75.6$ & $61.5$ & $\mathbf{47.2}$ & $85.8$ & $\mathbf{43.0}$ & $0.0$ & $44.4$ & $97.9$ & $\mathbf{0.0}$ & $78.1$ & $99.5$ & $\underline{18.1}$ & $70.7$ & $74.9$\\
  & DeepSeek-V4-Pro & $6.4$ & $\underline{85.5}$ & $92.0$ & $6.0$ & $\mathbf{93.9}$ & $93.5$ & $0.0$ & $\underline{65.3}$ & $100.0$ & $\mathbf{0.0}$ & $\mathbf{85.0}$ & $100.0$ & $3.1$ & $\mathbf{82.4}$ & $96.4$\\
\midrule
\multirow{8}{*}{\rotatebox[origin=c]{90}{\textbf{MCP}}} & GPT-OSS-20B & $15.2$ & $65.6$ & $68.4$ & $30.4$ & $77.0$ & $56.9$ & $0.0$ & $58.4$ & $\underline{93.5}$ & $\mathbf{0.0}$ & $65.1$ & $\underline{98.4}$ & $10.2$ & $62.0$ & $74.8$\\
  & Mistral-Small-3.2-24B-Instruct & $27.6$ & $70.6$ & $\underline{54.8}$ & $25.2$ & $77.1$ & $65.1$ & $\mathbf{2.3}$ & $24.1$ & $\mathbf{81.0}$ & $\mathbf{0.0}$ & $69.3$ & $\mathbf{98.0}$ & $13.6$ & $59.4$ & $\underline{73.8}$\\
  & Qwen3.6-27B & $25.1$ & $\underline{84.8}$ & $68.0$ & $12.1$ & $\mathbf{92.7}$ & $85.5$ & $0.0$ & $62.4$ & $100.0$ & $\mathbf{0.0}$ & $\underline{82.6}$ & $100.0$ & $9.3$ & $\underline{80.6}$ & $88.4$\\
  & Nemotron-3-Nano-30B-A3B & $10.8$ & $78.1$ & $81.5$ & $23.2$ & $87.6$ & $72.5$ & $0.0$ & $46.9$ & $95.5$ & $\mathbf{0.0}$ & $76.1$ & $99.5$ & $8.5$ & $72.1$ & $87.1$\\
  & Gemma-4-31B-it & $\underline{29.1}$ & $82.8$ & $62.0$ & $14.4$ & $87.4$ & $82.4$ & $0.0$ & $66.2$ & $100.0$ & $\mathbf{0.0}$ & $80.2$ & $99.5$ & $10.9$ & $79.0$ & $85.9$\\
  & GLM-5.2 & $\mathbf{68.6}$ & $79.6$ & $\mathbf{12.5}$ & $\underline{46.5}$ & $88.7$ & $\underline{47.5}$ & $\underline{1.9}$ & $\underline{66.8}$ & $96.4$ & $\mathbf{0.0}$ & $80.4$ & $100.0$ & $\mathbf{29.3}$ & $78.6$ & $\mathbf{63.8}$\\
  & Kimi-k2.6 & $23.1$ & $81.0$ & $69.5$ & $\mathbf{48.9}$ & $88.3$ & $\mathbf{44.5}$ & $0.0$ & $42.0$ & $99.5$ & $\mathbf{0.0}$ & $76.2$ & $100.0$ & $\underline{18.0}$ & $71.3$ & $77.0$\\
  & DeepSeek-V4-Pro & $7.6$ & $\mathbf{86.3}$ & $90.0$ & $5.9$ & $\underline{91.9}$ & $93.5$ & $0.0$ & $\mathbf{69.4}$ & $100.0$ & $\mathbf{0.0}$ & $\mathbf{85.0}$ & $100.0$ & $3.4$ & $\mathbf{83.1}$ & $95.8$\\
\midrule
\multirow{8}{*}{\rotatebox[origin=c]{90}{\textbf{OpenCode}}} & GPT-OSS-20B & $14.2$ & $45.4$ & $51.0$ & $39.6$ & $62.2$ & $\underline{33.0}$ & $\underline{10.5}$ & $42.5$ & $61.5$ & $\underline{0.0}$ & $37.1$ & $\mathbf{85.0}$ & $16.1$ & $46.8$ & $57.6$\\
  & Mistral-Small-3.2-24B-Instruct & $23.1$ & $61.6$ & $51.0$ & $35.8$ & $70.7$ & $47.5$ & $0.0$ & $0.0$ & $\mathbf{0.0}$ & $\underline{0.0}$ & $60.8$ & $\underline{94.5}$ & $14.7$ & $48.3$ & $\underline{48.2}$\\
  & Qwen3.6-27B & $\underline{53.2}$ & $\underline{81.0}$ & $\underline{34.5}$ & $23.2$ & $88.1$ & $73.0$ & $3.5$ & $64.3$ & $87.0$ & $\underline{0.0}$ & $\underline{81.3}$ & $100.0$ & $\underline{20.0}$ & $78.7$ & $73.6$\\
  & Nemotron-3-Nano-30B-A3B & $17.2$ & $69.2$ & $65.5$ & $\underline{51.3}$ & $75.0$ & $\mathbf{29.5}$ & $5.1$ & $34.2$ & $67.0$ & $\mathbf{0.1}$ & $63.5$ & $96.0$ & $18.4$ & $60.5$ & $64.5$\\
  & Gemma-4-31B-it & $19.3$ & $70.0$ & $66.5$ & $6.0$ & $75.5$ & $89.5$ & $3.0$ & $60.5$ & $81.5$ & $\underline{0.0}$ & $72.3$ & $99.0$ & $7.1$ & $69.6$ & $84.1$\\
  & GLM-5.2 & $\mathbf{75.2}$ & $79.5$ & $\mathbf{5.0}$ & $\mathbf{55.8}$ & $88.1$ & $36.5$ & $\mathbf{53.6}$ & $79.0$ & $\underline{34.0}$ & $\underline{0.0}$ & $80.1$ & $100.0$ & $\mathbf{46.2}$ & $\underline{81.7}$ & $\mathbf{43.9}$\\
  & Kimi-k2.6 & $18.7$ & $79.4$ & $71.5$ & $47.3$ & $\underline{88.4}$ & $47.0$ & $1.1$ & $\underline{80.5}$ & $97.0$ & $\underline{0.0}$ & $75.7$ & $99.5$ & $16.8$ & $81.0$ & $78.8$\\
  & DeepSeek-V4-Pro & $6.6$ & $\mathbf{85.5}$ & $90.0$ & $9.1$ & $\mathbf{93.1}$ & $90.5$ & $3.5$ & $\mathbf{84.8}$ & $94.5$ & $\underline{0.0}$ & $\mathbf{86.1}$ & $100.0$ & $4.8$ & $\mathbf{87.4}$ & $93.8$\\
\midrule
\multirow{8}{*}{\rotatebox[origin=c]{90}{\textbf{Hermes}}} & GPT-OSS-20B & $9.9$ & $24.1$ & $23.5$ & $8.6$ & $22.4$ & $\underline{23.5}$ & $\underline{8.8}$ & $57.5$ & $69.5$ & $\mathbf{0.3}$ & $27.4$ & $70.0$ & $6.9$ & $32.8$ & $46.6$\\
  & Mistral-Small-3.2-24B-Instruct & $7.5$ & $14.7$ & $8.5$ & $9.2$ & $17.2$ & $\mathbf{11.0}$ & $0.0$ & $0.0$ & $\mathbf{0.0}$ & $\underline{0.0}$ & $17.0$ & $\underline{25.5}$ & $4.2$ & $12.2$ & $\mathbf{11.2}$\\
  & Qwen3.6-27B & $\underline{68.9}$ & $73.0$ & $\underline{5.0}$ & $26.9$ & $79.3$ & $59.0$ & $\underline{8.8}$ & $58.8$ & $85.5$ & $\mathbf{0.3}$ & $77.8$ & $99.0$ & $\underline{26.2}$ & $72.2$ & $62.1$\\
  & Nemotron-3-Nano-30B-A3B & $10.9$ & $28.5$ & $34.5$ & $3.4$ & $9.5$ & $\mathbf{11.0}$ & $2.0$ & $\underline{62.9}$ & $95.5$ & $\underline{0.0}$ & $9.9$ & $\mathbf{21.5}$ & $4.1$ & $27.7$ & $40.6$\\
  & Gemma-4-31B-it & $37.5$ & $73.9$ & $42.5$ & $12.6$ & $84.8$ & $80.0$ & $5.5$ & $61.4$ & $82.0$ & $\underline{0.0}$ & $80.2$ & $99.0$ & $13.9$ & $\underline{75.1}$ & $75.9$\\
  & GLM-5.2 & $\mathbf{80.4}$ & $\underline{80.9}$ & $\mathbf{0.5}$ & $\mathbf{57.6}$ & $\mathbf{87.5}$ & $32.5$ & $\mathbf{62.1}$ & $\mathbf{76.0}$ & $\underline{17.5}$ & $\underline{0.0}$ & $\mathbf{82.4}$ & $100.0$ & $\mathbf{50.0}$ & $\mathbf{81.7}$ & $\underline{37.6}$\\
  & Kimi-k2.6 & $26.7$ & $62.4$ & $48.0$ & $\underline{36.1}$ & $68.2$ & $41.5$ & $2.6$ & $45.5$ & $84.0$ & $\underline{0.0}$ & $53.1$ & $83.0$ & $16.3$ & $57.3$ & $64.1$\\
  & DeepSeek-V4-Pro & $32.8$ & $\mathbf{82.5}$ & $58.5$ & $4.9$ & $\underline{86.9}$ & $91.0$ & $1.2$ & $46.1$ & $81.0$ & $\underline{0.0}$ & $\underline{82.2}$ & $99.5$ & $9.7$ & $74.4$ & $82.5$\\
\midrule
\multirow{8}{*}{\rotatebox[origin=c]{90}{\textbf{OpenClaw}}} & GPT-OSS-20B & $17.9$ & $42.1$ & $43.5$ & $28.8$ & $48.4$ & $39.0$ & $\underline{9.1}$ & $48.2$ & $69.0$ & $0.0$ & $45.0$ & $96.5$ & $14.0$ & $45.9$ & $62.0$\\
  & Mistral-Small-3.2-24B-Instruct & $22.9$ & $68.2$ & $59.5$ & $35.8$ & $73.8$ & $47.0$ & $0.0$ & $0.0$ & $\mathbf{0.0}$ & $\underline{0.1}$ & $63.7$ & $96.5$ & $14.7$ & $51.4$ & $50.8$\\
  & Qwen3.6-27B & $\underline{54.5}$ & $\underline{80.8}$ & $\underline{33.5}$ & $20.1$ & $\underline{85.6}$ & $74.5$ & $2.0$ & $41.6$ & $93.5$ & $0.0$ & $\mathbf{78.9}$ & $100.0$ & $\underline{19.2}$ & $\underline{71.7}$ & $75.4$\\
  & Nemotron-3-Nano-30B-A3B & $20.9$ & $44.0$ & $36.5$ & $\underline{47.4}$ & $50.1$ & $\mathbf{6.0}$ & $7.4$ & $11.6$ & $\underline{8.5}$ & $\mathbf{0.2}$ & $27.6$ & $\mathbf{76.5}$ & $19.0$ & $33.3$ & $\mathbf{31.9}$\\
  & Gemma-4-31B-it & $19.4$ & $71.0$ & $66.5$ & $10.1$ & $75.3$ & $85.5$ & $2.5$ & $\mathbf{65.2}$ & $93.5$ & $0.0$ & $71.1$ & $99.0$ & $8.0$ & $70.7$ & $86.1$\\
  & GLM-5.2 & $\mathbf{76.3}$ & $\mathbf{81.4}$ & $\mathbf{6.5}$ & $\mathbf{60.4}$ & $\mathbf{89.0}$ & $34.0$ & $\mathbf{32.1}$ & $\underline{60.2}$ & $43.0$ & $0.0$ & $\underline{78.2}$ & $97.0$ & $\mathbf{42.2}$ & $\mathbf{77.2}$ & $\underline{45.1}$\\
  & Kimi-k2.6 & $26.1$ & $60.3$ & $48.5$ & $26.2$ & $41.2$ & $\underline{29.0}$ & $2.5$ & $24.4$ & $49.0$ & $0.0$ & $55.3$ & $\underline{81.0}$ & $13.7$ & $45.3$ & $51.9$\\
  & DeepSeek-V4-Pro & $8.0$ & $78.0$ & $85.5$ & $11.8$ & $84.8$ & $85.5$ & $3.2$ & $41.9$ & $81.5$ & $0.0$ & $71.4$ & $98.0$ & $5.8$ & $69.0$ & $87.6$\\
\midrule
\multirow{8}{*}{\rotatebox[origin=c]{90}{\textbf{DeepSeek-Harness}}} & GPT-OSS-20B & $24.0$ & $61.9$ & $52.5$ & $34.1$ & $65.8$ & $44.5$ & $5.8$ & $48.1$ & $77.0$ & $\mathbf{0.0}$ & $36.8$ & $96.5$ & $\underline{16.0}$ & $53.1$ & $67.6$\\
  & Mistral-Small-3.2-24B-Instruct & $13.5$ & $46.7$ & $42.5$ & $24.7$ & $53.4$ & $\underline{35.0}$ & $0.0$ & $0.0$ & $\mathbf{0.0}$ & $\mathbf{0.0}$ & $47.3$ & $\underline{67.5}$ & $9.5$ & $36.9$ & $\mathbf{36.2}$\\
  & Qwen3.6-27B & $\underline{44.1}$ & $\underline{84.3}$ & $50.0$ & $13.2$ & $\mathbf{93.1}$ & $85.5$ & $0.0$ & $67.0$ & $100.0$ & $\mathbf{0.0}$ & $\mathbf{85.3}$ & $100.0$ & $14.3$ & $\mathbf{82.4}$ & $83.9$\\
  & Nemotron-3-Nano-30B-A3B & $13.7$ & $78.7$ & $78.5$ & $\underline{34.5}$ & $78.4$ & $55.0$ & $3.8$ & $59.0$ & $89.5$ & $\mathbf{0.0}$ & $75.8$ & $98.0$ & $13.0$ & $73.0$ & $80.2$\\
  & Gemma-4-31B-it & $13.9$ & $57.5$ & $56.5$ & $4.3$ & $58.7$ & $78.0$ & $0.2$ & $\underline{68.0}$ & $97.5$ & $\mathbf{0.0}$ & $40.0$ & $94.5$ & $4.6$ & $56.1$ & $81.6$\\
  & GLM-5.2 & $\mathbf{77.6}$ & $80.8$ & $\mathbf{3.5}$ & $\mathbf{43.5}$ & $89.0$ & $49.5$ & $\mathbf{39.0}$ & $\mathbf{69.2}$ & $37.5$ & $\mathbf{0.0}$ & $82.3$ & $99.5$ & $\mathbf{40.0}$ & $\underline{80.4}$ & $47.5$\\
  & Kimi-k2.6 & $6.7$ & $30.4$ & $\underline{32.5}$ & $22.9$ & $41.2$ & $\mathbf{26.0}$ & $0.2$ & $12.2$ & $\underline{29.0}$ & $\mathbf{0.0}$ & $46.5$ & $\mathbf{63.5}$ & $7.5$ & $32.6$ & $\underline{37.8}$\\
  & DeepSeek-V4-Pro & $6.1$ & $\mathbf{85.0}$ & $91.0$ & $10.5$ & $\underline{92.7}$ & $87.5$ & $\underline{7.8}$ & $35.2$ & $58.0$ & $\mathbf{0.0}$ & $\underline{85.2}$ & $99.5$ & $6.1$ & $74.6$ & $84.0$\\
\midrule
\multirow{8}{*}{\rotatebox[origin=c]{90}{\textbf{Pi}}} & GPT-OSS-20B & $18.5$ & $53.3$ & $55.5$ & $28.0$ & $62.6$ & $50.5$ & $\underline{10.3}$ & $41.3$ & $61.0$ & $\mathbf{0.0}$ & $55.6$ & $96.0$ & $14.2$ & $53.2$ & $65.8$\\
  & Mistral-Small-3.2-24B-Instruct & $30.1$ & $72.5$ & $52.0$ & $\underline{39.1}$ & $77.5$ & $46.5$ & $0.0$ & $0.0$ & $\mathbf{0.0}$ & $\mathbf{0.0}$ & $64.6$ & $98.0$ & $17.3$ & $53.7$ & $49.1$\\
  & Qwen3.6-27B & $\underline{54.2}$ & $\underline{84.1}$ & $36.0$ & $23.7$ & $\underline{92.4}$ & $75.0$ & $0.0$ & $\underline{73.4}$ & $98.5$ & $\mathbf{0.0}$ & $\underline{84.3}$ & $100.0$ & $\underline{19.5}$ & $\underline{83.6}$ & $77.4$\\
  & Nemotron-3-Nano-30B-A3B & $19.7$ & $74.6$ & $69.0$ & $31.4$ & $75.0$ & $50.5$ & $4.2$ & $52.3$ & $78.5$ & $\mathbf{0.0}$ & $64.1$ & $\underline{90.5}$ & $13.8$ & $66.5$ & $72.1$\\
  & Gemma-4-31B-it & $22.2$ & $68.6$ & $59.0$ & $6.0$ & $73.3$ & $87.0$ & $2.0$ & $69.4$ & $92.5$ & $\mathbf{0.0}$ & $67.9$ & $97.5$ & $7.5$ & $69.8$ & $84.0$\\
  & GLM-5.2 & $\mathbf{74.3}$ & $80.1$ & $\mathbf{7.0}$ & $\mathbf{51.3}$ & $88.8$ & $\underline{41.0}$ & $\mathbf{57.4}$ & $\mathbf{84.0}$ & $30.0$ & $\mathbf{0.0}$ & $82.1$ & $99.5$ & $\mathbf{45.7}$ & $\mathbf{83.7}$ & $\underline{44.4}$\\
  & Kimi-k2.6 & $13.8$ & $38.3$ & $\underline{29.0}$ & $20.9$ & $49.0$ & $\mathbf{32.0}$ & $0.0$ & $3.0$ & $\underline{7.5}$ & $\mathbf{0.0}$ & $38.4$ & $\mathbf{65.0}$ & $8.7$ & $32.2$ & $\mathbf{33.4}$\\
  & DeepSeek-V4-Pro & $4.8$ & $\mathbf{85.6}$ & $93.5$ & $9.2$ & $\mathbf{93.3}$ & $90.0$ & $1.6$ & $40.4$ & $76.5$ & $\mathbf{0.0}$ & $\mathbf{87.0}$ & $100.0$ & $3.9$ & $76.6$ & $90.0$\\
\bottomrule
\end{tabular}
\end{table*}

Because GLM-5.2 exhibits relatively low $ASR$ in Table~\ref{tab:rq2-attacks}, we use it as a conservative target to study whether the attack agent can improve difficult cases without changing the original attack goal. Concretely, we keep the attack goal fixed and optimize only the attack payload, so any improvement reflects better payload stealth and plausibility. The optimization process is illustrated in Figure~\ref{fig:attack-progress}. We observe that optimization substantially increases attack success for IPI, TPA, and SCI, showing that many initially unsuccessful attacks fail not because the goal is unattainable, but because the initial payload is not yet persuasive or robust enough for the target model and environment. This confirms that the attack agent's specification-guided optimization is effective, and that evaluating only one-shot attacks would underestimate the practical risk faced by deployed agents.

\begin{rqanswer}
\textbf{Answer to RQ3:} All four attacks are effective. TCI and SCI are the most consistently successful, while IPI and TPA are more model-sensitive. Moreover, even when GLM-5.2 initially shows relatively low $ASR$, optimization can still substantially improve success.
\end{rqanswer}

\subsection{Effectiveness of Existing Defenses (RQ4)}

We evaluate five defenses on OpenCode with DeepSeek-V4-Pro and GLM-5.2, focusing on whether any single method can cover all four attack surfaces. \textbf{AgentDoG} \cite{agentdog2026} and \textbf{ShieldAgent} \cite{AgentSafetyBench} are both guardrail models inserted before each tool call: the current context is formatted into their detection prompt, and if the model predicts the step as unsafe, the tool call is blocked and a refusal message is returned to the agent. \textbf{ClawGuard} \cite{clawguard2026} intercepts violating actions at the tool-call boundary based on defined rules and returns the blocking feedback to the agent so that the task can continue and the model may recover. In addition, we include two defenses designed by us to separate attack channels more clearly. \textbf{Scan \& Repair} runs a subagent before task execution to inspect tool code and skill code, repair suspicious hidden logic such as exfiltration behavior, and then execute the task using the repaired artifacts. \textbf{Prompt Detector} checks tool descriptions for injected instructions before the task starts and inserts a warning if suspicious content is found; during execution, after each tool return, it checks the returned content again and injects a warning message telling the agent which content may be malicious and should not be followed. Both Prompt Detector and Scan \& Repair use the same LLM as the main agent. Table \ref{tab:rq3-defenses} illustrates the defense effectiveness of these methods. AgentDoG and ShieldAgent provide only limited protection overall; a likely reason is that both are relatively small guardrail models trained on limited distributions, so their generalization to our more diverse and deceptive attack tasks is insufficient. ClawGuard is more effective on some text-mediated cases because it performs explicit dynamic blocking, but it still does not comprehensively cover all four attack surfaces, and stricter intervention can reduce utility. Our two attack-specific defenses behave as intended: Prompt Detector is more helpful for IPI and TPA, where the malicious instruction is visible in tool description or tool return, while Scan \& Repair is much more effective for SCI and TCI, where the harmful behavior is hidden inside code and therefore largely invisible to prompt-level defenses. 

\begin{table*}[t]
\centering
\caption{The defense effectiveness of different defense methods against different attacks.}
\label{tab:rq3-defenses}
\scriptsize
\setlength{\tabcolsep}{2.0pt}
\resizebox{\textwidth}{!}{%
\begin{tabular}{@{}cl*{5}{rrr}@{}}
\toprule
\multirow{2}{*}{\textbf{Model}} & \multirow{2}{*}{\textbf{Defense}} & \multicolumn{3}{c}{\textbf{IPI}} & \multicolumn{3}{c}{\textbf{TPA}} & \multicolumn{3}{c}{\textbf{SCI}} & \multicolumn{3}{c}{\textbf{TCI}} & \multicolumn{3}{c}{\textbf{Total}} \\
\cmidrule(lr){3-5}\cmidrule(lr){6-8}\cmidrule(lr){9-11}\cmidrule(lr){12-14}\cmidrule(lr){15-17}
& & $\mathbf{SCR}$ & $\mathbf{TCR}$ & $\mathbf{ASR}$ & $\mathbf{SCR}$ & $\mathbf{TCR}$ & $\mathbf{ASR}$ & $\mathbf{SCR}$ & $\mathbf{TCR}$ & $\mathbf{ASR}$ & $\mathbf{SCR}$ & $\mathbf{TCR}$ & $\mathbf{ASR}$ & $\mathbf{SCR}$ & $\mathbf{TCR}$ & $\mathbf{ASR}$ \\
\midrule
\multirow[c]{6}{*}{\rotatebox[origin=c]{90}{\tiny\textbf{DeepSeek-V4-Pro}}} & None & $6.60$ & $\mathbf{85.52}$ & $90.00$ & $\underline{9.13}$ & $\underline{93.09}$ & $\underline{90.50}$ & $3.50$ & $84.83$ & $94.50$ & $\underline{0.00}$ & $\mathbf{86.09}$ & $100.00$ & $4.81$ & $\underline{87.38}$ & $93.75$ \\
 & AgentDoG1.5-4B & $6.87$ & $84.37$ & $88.50$ & $8.00$ & $\mathbf{93.42}$ & $91.50$ & $3.67$ & $86.36$ & $94.00$ & $\underline{0.00}$ & $85.10$ & $\underline{99.50}$ & $4.63$ & $87.31$ & $93.38$ \\
 & ShieldAgent & $4.73$ & $\underline{85.15}$ & $93.00$ & $8.08$ & $92.39$ & $91.00$ & $5.42$ & $\underline{87.42}$ & $92.50$ & $\underline{0.00}$ & $85.30$ & $100.00$ & $4.56$ & $\mathbf{87.56}$ & $94.12$ \\
 & ClawGuard & $\underline{11.15}$ & $77.49$ & $\underline{80.00}$ & $4.71$ & $84.58$ & $92.50$ & $\underline{35.17}$ & $76.57$ & $\underline{54.00}$ & $\underline{0.00}$ & $76.80$ & $100.00$ & $12.76$ & $78.86$ & $81.62$ \\
 & Prompt Detector & $\mathbf{54.00}$ & $79.28$ & $\mathbf{29.50}$ & $\mathbf{34.50}$ & $90.11$ & $\mathbf{61.00}$ & $6.25$ & $81.83$ & $89.50$ & $\underline{0.00}$ & $59.97$ & $\underline{99.50}$ & $\underline{23.69}$ & $77.80$ & $\underline{69.88}$ \\
 & Scan \& Repair & $6.86$ & $83.38$ & $90.50$ & $6.88$ & $91.21$ & $91.00$ & $\mathbf{83.42}$ & $\mathbf{88.08}$ & $\mathbf{5.00}$ & $\mathbf{85.98}$ & $\underline{85.98}$ & $\mathbf{0.00}$ & $\mathbf{45.78}$ & $87.16$ & $\mathbf{46.62}$ \\
\midrule
\multirow[c]{6}{*}{\rotatebox[origin=c]{90}{\tiny\textbf{GLM-5.2}}} & None & $75.22$ & $79.46$ & $5.00$ & $55.79$ & $\underline{88.14}$ & $36.50$ & $53.62$ & $\mathbf{79.04}$ & $34.00$ & $\underline{0.00}$ & $80.05$ & $100.00$ & $46.16$ & $\mathbf{81.67}$ & $43.88$ \\
 & AgentDoG1.5-4B & $72.99$ & $78.91$ & $7.50$ & $46.45$ & $86.88$ & $45.00$ & $45.75$ & $70.33$ & $39.00$ & $\underline{0.00}$ & $\underline{80.22}$ & $99.50$ & $41.30$ & $79.08$ & $47.75$ \\
 & ShieldAgent & $\mathbf{75.45}$ & $\underline{80.25}$ & $6.00$ & $51.25$ & $86.76$ & $39.00$ & $45.46$ & $68.88$ & $33.00$ & $\underline{0.00}$ & $79.26$ & $\underline{99.00}$ & $43.04$ & $78.79$ & $44.25$ \\
 & ClawGuard & $74.30$ & $74.30$ & $\mathbf{0.50}$ & $\underline{56.65}$ & $78.12$ & $\underline{24.00}$ & $\mathbf{68.96}$ & $74.21$ & $\underline{9.50}$ & $\underline{0.00}$ & $71.53$ & $100.00$ & $\underline{49.98}$ & $74.54$ & $\underline{33.50}$ \\
 & Prompt Detector & $74.09$ & $76.68$ & $\underline{3.00}$ & $\mathbf{66.41}$ & $87.52$ & $\mathbf{23.50}$ & $45.58$ & $\underline{75.92}$ & $40.00$ & $\underline{0.00}$ & $58.37$ & $100.00$ & $46.52$ & $74.62$ & $41.62$ \\
 & Scan \& Repair & $\underline{75.38}$ & $\mathbf{82.62}$ & $8.50$ & $56.40$ & $\mathbf{88.66}$ & $37.00$ & $\underline{63.88}$ & $67.04$ & $\mathbf{3.50}$ & $\mathbf{83.64}$ & $\mathbf{85.14}$ & $\mathbf{1.50}$ & $\mathbf{69.83}$ & $\underline{80.87}$ & $\mathbf{12.62}$ \\
\bottomrule
\end{tabular}}
\end{table*}

\begin{rqanswer}
\textbf{Answer to RQ4:} No evaluated defense can comprehensively defend against all four attack surfaces.  Therefore, comprehensive defense for LLM agents requires layered mechanisms that jointly protect untrusted text inputs, hidden executable artifacts, and dangerous actions.
\end{rqanswer}

\subsection{Quality of Generated Specifications (RQ5)}
\label{sec:rq5}
Automatic specification generation is still a challenging problem \cite{li2024formalllm,wang2025agentspec,mao2025llm,liu2026agents4plc}. In fact, even for human experts, writing complete and perfectly precise utility specifications is difficult, because realistic tasks often admit multiple valid workflows, implicit success conditions, and environment-dependent contingencies. We therefore do not expect automatically generated specifications to be perfect, but instead evaluate whether they are sufficiently faithful for practical benchmarking. As shown in Table~\ref{tab:rq4-agentdojo}, we compare the generated specifications against AgentDojo's rule-based evaluators~\cite{debenedetti2024agentdojo}, using AgentDojo's outputs as reference labels for both utility and security. The results show strong agreement overall, especially on security, while utility agreement is somewhat lower. This pattern is consistent with the fact that security goals are usually narrower and easier to formalize than task-completion conditions. We further manually inspected disagreement cases and found two main causes. First, many AgentDojo tasks require the LLM's final natural-language answer to contain certain specific strings, whereas our specifications usually avoid such rigid output-surface constraints and focus instead on whether the required actions and environment effects actually occurred. Second, our annotation agent sometimes generates utility specifications that unnecessarily constrain one particular execution path, thereby rejecting alternative but still acceptable solutions; this issue can be mitigated by the reflection step. 

\begin{table}[!t]
\centering
\begin{minipage}[!t]{0.5\linewidth}
\centering
\caption{Agreement between specification-based evaluation and AgentDojo's evaluators using AgentDojo's outputs as labels, including an ablation without the annotation agent's reflection.}
\label{tab:rq4-agentdojo}
\scriptsize
\begin{tabular}{cccccc}
\toprule
\textbf{Module} & \textbf{Dimension} & \textbf{Acc.} & \textbf{Prec.} & \textbf{Rec.} & \textbf{F1} \\
\midrule
\multirow{2}{*}{annotation agent} & utility & $84.5$ & $92.9$ & $89.8$ & $91.3$ \\
 & security & $95.2$ & $97.3$ & $91.6$ & $94.4$ \\
\hline
\multirow{2}{*}{w/o reflection} & utility & $80.4$ & $92.6$ & $85.2$ & $88.8$ \\
 & security & $93.5$ & $92.4$ & $92.7$ & $92.6$ \\
\bottomrule
\end{tabular}
\end{minipage}
\hfill
\begin{minipage}[!t]{0.45\linewidth}
\centering
\caption{Time (s) and space (bytes) overhead of specification-based evaluation, comparing with LLM-as-a-Judge and Agent-as-a-Judge in Section \ref{sec:rq1}. \#Events is the average events number.}
\label{tab:rq5-verification-efficiency}
\scriptsize
\begin{tabular}{cccccc}
\toprule
\textbf{Dataset} & \textbf{\#Events} & \textbf{Bytes} & $T_{\text{spec}}$ & $T_{\text{LLM}}$ & $T_{\text{Agent}}$ \\
\midrule
IPI & $22.0$ & $809$ & $0.3025$ & $55.35$ & $92.80$ \\
TPA & $21.9$ & $802$ & $0.3007$ & $46.87$ & $90.62$ \\
SCI & $36.5$ & $440$ & $0.3097$ & $36.65$ & $97.36$ \\
TCI & $25.9$ & $768$ & $0.3046$ & $59.28$ & $104.1$ \\
\bottomrule
\end{tabular}
\end{minipage}
\end{table}

\begin{rqanswer}
\textbf{Answer to RQ5:} Automatic specification generation is feasible but still imperfect, especially for utility properties. Reflection helps reduce unnecessary path constraints, but some cases still require human review.
\end{rqanswer}

\subsection{Efficiency of Specification-based Evaluation (RQ6)}

We evaluate the efficiency of RAS-Eval on the final trajectories produced by DeepSeek-V4-Pro with OpenCode. For each execution, we compile it into a nuXmv model and measure the verification cost. We also compare our verification time with the time required by LLM-as-a-Judge and Agent-as-a-Judge in Section \ref{sec:rq1} under the same hardware configuration. In addition, we record the size of the generated nuXmv models. Table~\ref{tab:rq5-verification-efficiency} illustrates the results. The actual operation requires verification of each step, so the average overall verification time for each task does not exceed $\text{\#Events} \times T_{\text{LTL}} $. These results show that our method is substantially more efficient than both LLM-as-a-Judge and Agent-as-a-Judge, because verification does not require additional LLM inference or agent-side environment exploration. Furthermore, the generated nuXmv models remain compact, indicating that the predicate-abstraction-based encoding introduces very low storage overhead.

\begin{rqanswer}
\textbf{Answer to RQ6:} Specification-based evaluation has low overhead in time compared to LLM-as-a-Judge and Agent-as-a-Judge, making RAS-Eval practical for large-scale security evaluation.
\end{rqanswer}

\section{Related Work}

\textit{LLM Agent Security Evaluation.} Security evaluation for tool-using LLM agents has progressed from prompt-level attacks to executable, system-level benchmarks. Early work formalizes prompt injection and demonstrates indirect prompt injection in LLM-integrated applications~\cite{IPI,liu2024formalizing,injecagent}. AgentDojo evaluates attacks and defenses in a stateful tool environment, while ToolEmu scales risk discovery through LM-emulated tools~\cite{debenedetti2024agentdojo,ruan2024toolemu}. Broader suites such as ASB and Agent-SafetyBench cover multiple attack or risk categories~\cite{asb,AgentSafetyBench}, whereas R-Judge and AgentAuditor study learned risk judgment over trajectories~\cite{rjudge,luo2025agentauditor}. Recent work further exposes environment-, skill-, and MCP-specific attack surfaces through EIA, Skill-Inject, MCPTox and MSB~\cite{liao2024eia,skill_inject,msb, mcptox}. Complementary frameworks define contextual security properties and advocate monitoring artifacts across the full application stack~\cite{siu2026agentsecurity,zhang2026systemmonitoring}. Unlike benchmark-specific success scripts, simulated environments, or learned judges, RAS-Eval provides a shared deterministic oracle over tool-call trajectories and dynamic environment states.

\textit{Formal Methods for LLM Agents.} Recent work has begun to embed formal mechanisms around agent execution to improve the reliability and safety of LLM-based systems~\cite{zhang2025trustworthy}. AgentSpec monitors and intervenes in agent behavior during runtime based on formal specifications~\cite{wang2025agentspec}. AIR and ProbGuard further extend this direction with stronger semantic reasoning and probabilistic specification capabilities~\cite{xiao2026air,probguard}. Progent provides a rule-based programmable permission control framework \cite{progent}. AgentArmor tracks and controls the information flow during the operation of agents by combining LLM perception and rule checking \cite{agentarmor}. These systems provide important specification and enforcement capabilities, but mainly target defensive runtime monitoring and intervention. In contrast, RAS-Eval is designed as a deterministic evaluation oracle: it uses a more expressive temporal language over tool-call trajectories and dynamic environment observations to assess whether security and utility properties were satisfied in a realized execution across heterogeneous agent harnesses.

\section{Limitations and Discussion}
RAS-Eval still has several limitations. Specifications are not perfect, and automatically generating perfect specifications for realistic tasks remains unsolved. Security specifications only capture attack-task-related forbidden effects rather than arbitrary risky behaviors, but this is also a common boundary of existing agent benchmarks, which can only check predefined goals. In addition, our evaluation is designed to run with the agent; when only an existing trajectory is available, direct evaluation may require rerunning the agent. To reduce this cost, we provide a replay mechanism that re-executes recorded tool calls step by step to restore and observe intermediate environment states without calling the LLM again. However, some real tools are nondeterministic, so replay may rarely differ from the original run, which is ultimately a limitation of the tools themselves and the imperfect quality of the specifications.

\section{Conclusion}
We presented RAS-Eval, a specification-based framework for evaluating tool-using LLM agents through LTL-based formal verification over tool-call trajectories and dynamic environment observations. We also constructed a benchmark with $200$ normal tasks and $800$ attack tasks covering IPI, TPA, SCI, and TCI. RAS-Eval provides more stable evaluation, captures transient violations missed by final-state-only checking, and reveals that no evaluated defense comprehensively covers all four attack surfaces. These results suggest that formal verification is a practical complement to existing LLM-based and rule-based methods for security evaluation of LLM agents.

\bibliographystyle{ACM-Reference-Format}
\bibliography{refs}

\end{document}